\documentclass[aps,prx,showpacs,amsmath,amssymb,superscriptaddress,twocolumn,longbibliography,10pt]{revtex4-2}
\usepackage[english]{babel}
\usepackage[utf8]{inputenc}
\usepackage{amsfonts}
\usepackage[T1]{fontenc}
\usepackage[pdftex]{graphicx}
\usepackage{url}
\usepackage[titletoc,title]{appendix}
\usepackage{epstopdf}
\usepackage{amsmath} 
\usepackage{amssymb}
\usepackage{dsfont}
\usepackage{braket}
\usepackage{mathtools}
\usepackage{xcolor}
\usepackage{tikz}
\usepackage[export]{adjustbox}
\usetikzlibrary{arrows.meta}
\usepackage{hyperref}
\usepackage{bbold}
\usepackage[normalem]{ulem}
\usepackage{adjustbox}
\hypersetup{
    unicode=false, 
    pdftoolbar=false, 
    pdfmenubar=true, 
    pdffitwindow=false, 
    pdfstartview={}, 
    pdftitle={Finding periodic orbits in projected quantum many-body dynamics}, 
    pdfauthor={E. Petrova et al.}, 
    pdfsubject={}, 
    pdfcreator={}, 
    pdfproducer={}, 
    pdfkeywords={MPS, tangent space, time evolution, periodic orbits}, 
    pdfnewwindow=true, 
    colorlinks=true, 
    linkcolor=black, 
    citecolor=black, 
    filecolor=black, 
    urlcolor=black 
}
\usepackage{amsmath}
\usepackage{algpseudocodex}
\usepackage{newfloat}
\DeclareFloatingEnvironment[name=Algorithm]{algorithm}  
\newcommand{\algorule}{\noindent\rule{\linewidth}{1pt}}

\usepackage{outline}

\makeatletter
\newcommand{\ostar}{\mathbin{\mathpalette\make@circled\star}}
\newcommand{\make@circled}[2]{
  \ooalign{$\m@th#1\smallbigcirc{#1}$\cr\hidewidth$\m@th#1#2$\hidewidth\cr}
}
\newcommand{\smallbigcirc}[1]{
  \vcenter{\hbox{\scalebox{0.77778}{$\m@th#1\bigcirc$}}}
}
\makeatother

\begin{document}
\title{Finding periodic orbits in projected quantum many-body dynamics} 
\author{Elena Petrova}
\affiliation{Institute of Science and Technology Austria (ISTA), Am Campus 1, 3400 Klosterneuburg, Austria}
\author{Marko Ljubotina}
\affiliation{Institute of Science and Technology Austria (ISTA), Am Campus 1, 3400 Klosterneuburg, Austria}
\affiliation{Physics Department, Technical University of Munich, TUM School of Natural Sciences, Lichtenbergstr. 4,
Garching 85748, Germany}
\affiliation{Munich Center for Quantum Science and Technology (MCQST), Schellingstr. 4, München 80799, Germany}
\author{G\"okhan Yaln{\i}z}
\affiliation{Institute of Science and Technology Austria (ISTA), Am Campus 1, 3400 Klosterneuburg, Austria}
\author{Maksym Serbyn}
\affiliation{Institute of Science and Technology Austria (ISTA), Am Campus 1, 3400 Klosterneuburg, Austria}
    \begin{abstract}
        Describing general quantum many-body dynamics is a challenging task due to the exponential growth of the Hilbert space with system size. The time-dependent variational principle (TDVP) provides a powerful tool to tackle this task by projecting quantum evolution onto a classical dynamical system within a variational manifold. In classical systems, periodic orbits play a crucial role in understanding the structure of the phase space and the long-term behavior of the system. However, finding periodic orbits is generally difficult, and their existence and properties in generic TDVP dynamics over matrix product states have remained largely unexplored. In this work, we develop an algorithm to systematically identify and characterize periodic orbits in TDVP dynamics. Applying our method to the periodically kicked Ising model, we uncover both stable and unstable periodic orbits. We characterize the Kolmogorov-Arnold-Moser tori in the vicinity of stable periodic orbits and track the change of the periodic orbits as we modify the Hamiltonian parameters. We observe that periodic orbits exist at any value of the coupling constant between prethermal and fully thermalizing regimes, but their relevance to quantum dynamics and imprint on quantum eigenstates diminishes as the system leaves the prethermal regime. Our results demonstrate that periodic orbits provide valuable insights into the TDVP approximation of quantum many-body evolution and establish a closer connection between quantum and classical chaos.   
    \end{abstract}
    \maketitle
    
    \section{Introduction}

The study of quantum chaos has traditionally focused on few-body quantum systems with well-defined semiclassical limits, drawing parallels between classical and quantum dynamics \cite{haake1991quantum}. Seminal works have established criteria for quantum signatures of chaos, such as level statistics \cite{bohigas1984characterization, berry1977level} and out-of-time ordered correlators, in systems with an underlying classical counterpart \cite{larkin1969quasiclassical, swingle2016measuring, huang2017out, rozenbaum2017lyapunov}. However, recent advances in quantum simulation have shifted attention toward discrete quantum many-body systems, such as spin-1/2 chains, which, at first glance, are far from any semiclassical description.

Although far from the usual semiclassical limit, generic quantum models with small local Hilbert spaces are still expected to be chaotic, as is often probed by their level statistics \cite{oganesyan2007localization,d2016quantum, santos2010onset}. Typical dynamics in such models lead to rapid equilibration, where the system forgets the details of the initial state, relaxing to a state determined only by values of globally conserved quantities, such as energy -- a process known as thermalization. 
Despite recent progress in experiments  \cite{kaufman2016quantum, neill2016ergodic, scherg2021observing, bluvstein2021controlling,morong2021observation} and numerical studies \cite{lerose2021scaling, bukov2015prethermal, schindler2024geometric, michailidis2020slow}, understanding chaos and thermalization in discrete interacting quantum systems remains an open and active area of research.
 
To investigate the dynamics of these many-body quantum systems, matrix product states (MPS) have emerged as a powerful numerical tool \cite{doria2011optimal, cecile2024measurement}. Unlike the manifold of all possible product states, MPS systematically incorporate short-range entanglement, making them a more expressive variational class. Several methods exist for approximating full unitary quantum dynamics using MPS \cite{vidal2003efficient, vidal2004efficient, vidal2007classical}, with the time-dependent variational principle (TDVP) \cite{dirac1930note, lubich2008quantum, frenkel1934wave, haegeman2011time} standing out due to its special properties. 

It was already observed by P. Dirac that TDVP naturally leads to a symplectic classical dynamical system~\cite{dirac1930note}. When TDVP is applied to project dynamics onto the manifold of MPS, it provides a unique perspective: the quantum unitary evolution can be mapped onto an effective classical system that includes additional entanglement-related degrees of freedom, with the bond dimension (rank) of MPS $\chi$ controlling the amount of entanglement. This perspective provides a bridge between classical and quantum chaos, where tools from dynamical systems theory can be applied to study quantum thermalization, integrability, and emergent classical-like structures within quantum dynamics.

The emergence of classical dynamical systems from TDVP projections onto the MPS manifold motivates a systematic exploration of their properties \cite{michailidis2020slow,hallam2019lyapunov,ho2019periodic}. However, these systems are typically high-dimensional: for a spin-$1/2$ system, TDVP, even with the simplest nontrivial bond dimensions $\chi=2,3,\ldots$, results in a classical system with $\chi^2=4, 9,\ldots$ pairs of canonical coordinates and momenta. Classical chaos theory suggests that periodic orbits serve as fundamental structures that organize the phase space~\cite{cvitanovic2005chaos,Artuso90p1,Artuso90p2,cvitanovic91} of a system, providing crucial insights into stability, transport, and ergodicity~\cite{Eckmann85}. Despite their importance, periodic orbits in generic TDVP-MPS dynamics have remained largely unexplored due to the challenges of identifying them in high-dimensional systems, where symmetries and gauge invariance complicate the navigation of the phase space for the search for periodic orbits~\cite{cvitanovic2005chaos}.

In this work, we formulate and implement a general algorithm to search for periodic orbits within the MPS variational manifold. Using the kicked Ising model \cite{akila2016particle, bertini2018exact, bertini2019exact, bertini2019entanglement, piroli2020exact, lerose2021influence,d2013many, d2016quantum, lazarides2014equilibrium, mori2016rigorous, mori2018thermalization, neyenhuis2017observation, ho2023quantum, schindler2024geometric, ikeda2024robust} — a paradigmatic model for studying quantum thermalization — as an example, we identify periodic orbits and classify them in terms of their stability by computing their Floquet multipliers. For stable orbits, we visualize the surrounding Kolmogorov-Arnold-Moser (KAM) tori \cite{kolmogorov1954conservation, moser1962invariant, arnold2009proof, ott2002chaos} and show that their dimensionality scales as $\chi^2$, in agreement with expectations. Then, we track the change of these periodic orbits as system parameters tune the kicked Ising model from a prethermal regime \cite{fleckenstein2021thermalization} to a maximally chaotic dual-unitary point \cite{akila2016particle, bertini2018exact, bertini2019exact, bertini2019entanglement, piroli2020exact}. In the prethermal regime, we demonstrate that our approach can find variational approximations to eigenstates of the unitary operator generating dynamics over one period in the thermodynamic limit. As the system approaches the dual-unitary point, periodic orbits do not vanish but become unstable, shifting towards highly entangled regions of the MPS manifold, where entanglement reaches its maximal allowed value given the bond dimension, and leakage is typically higher.

Our findings establish that periodic orbits are present in TDVP-based projections of quantum many-body dynamics onto MPS manifolds. These orbits, both stable and unstable, offer a new perspective on quantum thermalization, providing direct links between quantum and classical chaos. Understanding their structure with increasing bond dimension offers a promising route to uncovering deeper connections between integrability, chaos, and entanglement in quantum many-body systems.

\section{Method for periodic orbits search}
\label{sec:II}

In this section, we discuss the building blocks of the algorithm used to find periodic orbits. First, we introduce the matrix product state (MPS) representation of the wave function and the time-dependent variational principle (TDVP) that we use to project the dynamics of quantum systems onto the MPS manifold. After this, in Sec.~\ref{sec:GD} we explain the main idea behind the algorithm for periodic orbit search, delegating the detailed description to Appendix~\ref{App:1}.
    
\subsection{iMPS and the time-dependent variational principle.}
\label{sec:iMPS}

The MPS \cite{vanderstraeten2019tangent} representation of a wave function may be viewed as a systematic improvement of uncorrelated product states that introduces entanglement. In the MPS ansatz, the state is expressed as a contraction of tensors 
\begin{equation}
\adjincludegraphics[width=0.3\linewidth,valign=c,raise=-3mm]{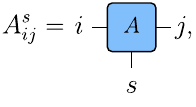}~
    \label{fig:A-def}
\end{equation}
with bond indices $i,j$ having bond dimensions $\chi_1$ and $\chi_2$, and the physical index $s$ running over $d$ values corresponding to the local Hilbert space dimension. We fix $d=2$, corresponding to spin-1/2, for the rest of this work. The contraction in Eq.~\eqref{fig:A-def} is performed over bond indices, which allows for the systematic introduction of correlations between different physical sites. For finite systems, each physical degree of freedom is assigned a different tensor within the MPS ansatz, potentially making translational symmetry difficult to identify. In contrast, infinite MPS (iMPS) \cite{vidal2007classical, haegeman2011time} explicitly resolve translational invariance by employing a single tensor $\chi\times d \times \chi$ that repeats infinitely, thus describing translationally invariant states in the thermodynamic limit:
\begin{equation}
    \begin{aligned}
      &\text{\raisebox{-5mm}{\includegraphics[width=0.85\linewidth]{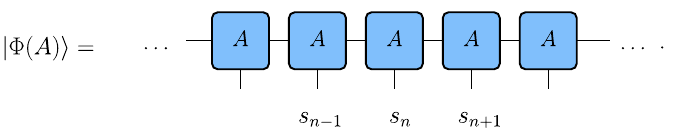}}}
    \end{aligned}
    \label{fig:AAA}
\end{equation}
        
The MPS representation of a wave function is not unique, as different choices of tensors $A^s_{ij}$ can still describe the same state. This non-uniqueness is known as the gauge freedom of MPS. To partially fix the gauge in Eq.~(\ref{fig:AAA}), we use the mixed canonical form of iMPS, which is given by: 
\begin{equation}
    \vcenter{\hbox{
        \includegraphics[width=0.88\linewidth, valign=c, raise=-3mm]{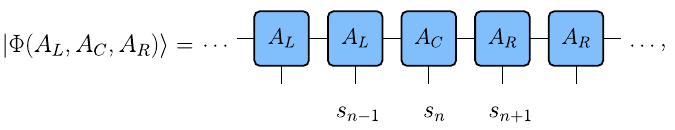}
        \label{fig:LCR}
    }}
    \end{equation}
where tensors $A_L$ and $A_R$ obey the orthogonality conditions:
\begin{equation}
    \begin{aligned}
      &\text{\raisebox{-10mm}{\includegraphics[width=0.81\linewidth]{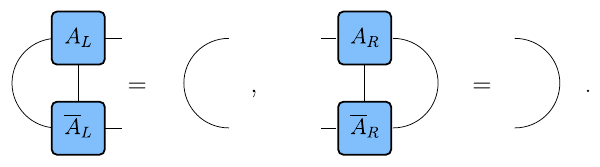}}}
    \end{aligned}
    \label{Eq:ALR}
\end{equation}

Tensors $A_L$, $A_C$, and $A_R$ are obtained from the one tensor $A$ from Eq.~\eqref{fig:AAA} and both forms, Eq.~\eqref{fig:AAA} and Eq.~\eqref{fig:LCR}, represent the same physical state. Another tensor that we are going to use in this work is the tensor $C$, which is defined as
$(A_L)_{ij}^sC_{jk} = (A_C)^s_{ik}$.
Na\"ively, the number of variables of an iMPS with a given bond dimension $\chi$ corresponds to the total number of entries in the tensor, that is $d \chi^2$ complex numbers. However, due to gauge transformations that leave the physical state invariant, not all of these parameters are physically meaningful. After accounting for this redundancy, for example, by imposing the left or right canonical conditions, the number of complex-valued variables describing the iMPS reduces to $(d-1) \chi^2$.

Another key concept relevant to our work is the iMPS tangent space. First of all, the tangent space is used in the TDVP algorithm to perform the time evolution. Second, we are using a tangent space basis in our gradient descent algorithm to compute gradients. 
The tangent space for iMPS is defined as the set of all possible infinitesimal changes of the iMPS that correspond to physical changes of the quantum state. 
The most general form of the tangent vector has the following form:
\begin{equation}
\begin{aligned}
  &\text{\raisebox{-1mm}{\includegraphics[width=0.9\linewidth]{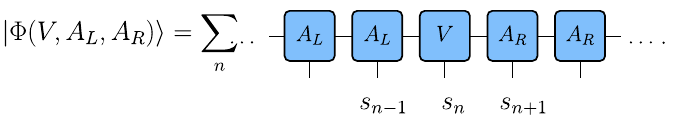}}}
\end{aligned}
\label{fig:LVR2}
\end{equation}
where tensor $V$ parametrizes directions in the tangent space. We leave the detailed discussion of the construction of the tangent space to Appendix~\ref{App:TDVP}. Here we only introduce the tangent basis as two sets of tensors $\{V^{(\alpha)}_{\text{re}}\}$ and $\{V^{(\alpha)}_{\text{im}}\}$, which satisfy the relation:
\begin{equation}
     i V^{(\alpha)}_{\text{re}}  =  V^{(\alpha)}_{\text{im}} ,
    \label{eq:basis}
\end{equation}
where index $\alpha$ runs over $\alpha \in [1, (d-1)\chi^2]$.

To simulate the dynamics of the quantum state, we employ the TDVP algorithm for iMPS in the mixed gauge. The core idea of this method is to project the action of the Hamiltonian $H$ onto the MPS tangent space, thus ensuring that the state evolution remains within the MPS manifold.
This key idea behind this method comes from the fact that directly applying the Hamiltonian to a state drives it out of the MPS manifold, and TDVP performs the orthogonal projection back to the MPS tangent space. By making an orthogonal projection to the MPS tangent space at each step, TDVP provides an approximation of the full quantum evolution restricted to the variational manifold. 
The TDVP equation for the quantum system is derived from the projected Schrödinger equation:
\begin{equation} \label{Eq:TDVP-derive}
    \partial_t |\psi(A_L, A_C, A_R)\rangle = -i\mathcal{P}H|\psi(A_L,A_C, A_R)\rangle,
\end{equation}
where $\mathcal{P}$ here projects the action of the Hamiltonian on the state to the tangent space manifold, approximating the derivative of the full Schrödinger equation. 

The presence of the projector onto the MPS tangent space in Eq.~(\ref{Eq:TDVP-derive}) above transforms the otherwise linear Schödinger equation into a highly non-linear differential equation. In terms of iMPS tensors, these equations of motion can be expressed as follows:
\begin{subequations} \label{Eq:EOMs}
\begin{eqnarray}
        \dot{A}_C &=& -iG_1(A_C),\\
        \dot{C} &=& iG_2(C),
\end{eqnarray}
\end{subequations}
where $G_1 $ and $G_2$ are linear Hermitian maps that implement the action of the projected Hamiltonian on the state. Although, as written, the equations for $A_C$ and $C$ look decoupled, in reality, they are interdependent. Indeed, as we discuss in Appendix~\ref{App:TDVP} (see also reviews~\cite{vanderstraeten2019tangent, zauner2018variational}), the linear Hermitian maps depend on tensors $A_{L,R}$, which can be obtained only from knowing both $A_C$ and $C$. We note that the TDVP projection of unitary dynamics leads to symplectic classical dynamics, meaning it preserves the symplectic structure of phase space, ensuring the conservation of fundamental geometric properties \cite{wu2020time}. In particular, the TDVP dynamics preserve the phase volume. While the presence of symplectic structure in equations of motion~(\ref{Eq:EOMs}) is not apparent, its consequences will be demonstrated below.

Since TDVP projects the true quantum dynamics onto the MPS manifold, some information inevitably gets lost over long-time evolution. This is related to the instantaneous leakage \cite{michailidis2020slow,ho2019periodic}, which is defined as 
\begin{equation}\label{eq:leak}
\gamma^2(t) = |||\dot{\psi} \rangle + iH|\psi\rangle ||^2 = ||\mathcal{P} H |\psi \rangle - H|\psi\rangle ||^2,
\end{equation}
where $|\psi\rangle$ is a state in the MPS manifold.
The leakage is fundamentally linked to entanglement growth as an MPS with a fixed bond dimension $\chi$ can only accurately represent states with limited entanglement, which is upper-bounded by $S_{\rm{ent}} \leq \ln \chi$ (here and below, entanglement is defined as the von Neumann entanglement entropy of the reduced density matrix $\rho_{1/2}$ of half of the system: $S_{\rm{ent}} =-\mathop {\rm tr} \rho_{1/2}\ln \rho_{1/2}$). However, during real-time dynamics, the entanglement entropy of a generic interacting quantum system is expected to grow linearly in time~\cite{KimHuse13}. Thus, as exact unitary dynamics would tend to depart from the MPS manifold of a fixed bond dimension -- a phenomenon quantified by the leakage -- this also leads to additional entanglement growth compared to that predicted by TDVP. Therefore, in this work, we compute the entanglement entropy from numerically exact time evolution and compare it with TDVP predictions to understand how leaky trajectories are. 

\subsection{Gradient descent method for periodic orbits}
\label{sec:GD}
To find periodic orbits, we need a reliable way to compare matrix product states to determine their similarity.
For this purpose, we introduce a measure of distance between two states (fidelity) using the mixed transfer matrix defined for two iMPS states specified by tensors $A_i$ and $A_j$ as follows:
\begin{equation}
    \begin{aligned}
      &\text{\raisebox{-5mm}{\includegraphics[width=0.4\linewidth]{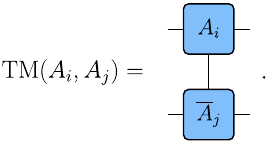}}}
    \end{aligned}
     \label{fig:TM}
\end{equation}
Merging the left and right indices allows us to represent ${\rm TM}(A_i, A_j)$ as a $\chi^2\times \chi^2$ matrix and study its spectra. 
In what follows, we focus on injective normalized MPS 
states~\cite{MPSRepresentations,MPSReview} which have a \emph{unique dominant eigenvalue} of the transfer matrix ${\rm TM}(A,A)$ with magnitude one~\footnote{We believe that our method can be extended to non-injective MPS, which are relevant in scenarios involving spontaneous symmetry breaking. However, a detailed investigation of this case is left for future work.}. Having two such states specified by tensors $A_i$ and $A_j$, we define the fidelity density (fidelity per site, hereafter referred to simply as fidelity) as:
\begin{equation}
            \label{eq:distance}
            \mathcal{D}(A_i,A_j) = \rho[\text{TM}(A_i,A_j)],
\end{equation}
where $\rho[\cdot]$ denotes the spectral radius, i.e., the largest absolute value among the eigenvalues of the transfer matrix.
This function is invariant with respect to MPS gauge transformations. Nevertheless, we will be using a left-canonical form of tensors in what follows. When $ \mathcal{D}(A_i, A_j) = 1$, the two MPS ($A_i$ and $A_j$) represent the same state in the thermodynamic limit. 

To find periodic orbits, we want to maximize the fidelity between the initial and time-evolved states. 
The evolved state is obtained as $A_L(T) = U^\text{TDVP}_T(A_L)$, where the $U^\text{TDVP}_T(\cdot)$ is the TDVP time propagation to time $T$ according to Eq.~\eqref{Eq:TDVP-derive}. Using the fidelity density introduced above, we define the cost function as the fidelity between the initial state and its time-evolved counterpart: 
\begin{equation}
    F(A_L) =  \mathcal{D}(U^\text{TDVP}_T(A_L), A_L).
    \label{eq:F}
\end{equation}
In this notation, finding a closed orbit is equivalent to finding an initial tensor $A_L(0)$ such that $F(A_L(0)) = 1$.
A naïve approach to finding the periodic orbit would be to apply a standard gradient-based optimization algorithm to maximize the cost function in Eq.~\eqref{eq:F}. However, the MPS has redundancies due to the gauge freedom, and such a naïve approach will also calculate gradients corresponding to the gauge directions that do not change the quantum state. This results in an additional overhead (e.g. for spin-1/2 systems, the MPS has $4\chi^2$ real parameters, however, the tangent space has only $2\chi^2$ independent vectors),
and also leads to spurious flat directions in the optimization landscape. 

Therefore, in our algorithm, we maximize the cost function~(\ref{eq:F}) by restricting the gradients to the MPS tangent space basis, improving computational efficiency and avoiding the unphysical directions. We use the tangent space basis construction discussed in Sec.~\ref{sec:iMPS} and apply the central difference method to compute the gradient. We combine the gradient components into an update term $\delta A_L$ summing each gradient component with the corresponding basis vector. The overall scheme for calculating an update for the tensor $A_L$, $\delta A_L$, is the following:
\begin{equation}
\label{eq:grads}
    \begin{gathered}
        g_{\text{re},\text{im}}^{\alpha} = \frac{F(A_L + \Delta V_{\text{re},\text{im}}^{(\alpha)}) - 
        F(A_L - \Delta V_{\text{re},\text{im}}^{(\alpha)})}{2\Delta},\\
        \delta A_L
        = \sum_{\alpha} (g_{\text{re}}^{\alpha} + ig_{\text{im}}^{\alpha})V^{(\alpha)}_{\text{re}},
    \end{gathered}
\end{equation} 
where $V^{(\alpha)}_{\text{re},\text{im}}$ are the tangent space basis vectors from Sec.~\ref{sec:iMPS}, and we used Eq.~\eqref{eq:basis} to simplify the second equation. 
Here we work only with the left canonical ($A_L$) form, as we use the left gauge fixing for the tangent basis, but a similar procedure is also possible for the right canonical form and corresponding tangent space basis.

In Eq.~\eqref{eq:grads} $\delta A_L$ plays the role of direction along which the tensor $A_L$ should be updated. To ensure numerical stability, we normalize $\delta A_L$ by dividing it by the square root of the largest eigenvalue of its transfer matrix $\text{TM}(\delta A_L, \delta A_L)$.
Then the tensor $A_L$ is updated by the normalized $\delta \tilde{A}_L$  times the learning rate $c$,
\begin{eqnarray}
    \tilde{A}_L = A_L + c \delta \tilde{A}_L.
\end{eqnarray}
The learning rate $c$ is adjusted adaptively during optimization: to mitigate slow convergence in flat regions, whenever the learning rate leads to an improvement in the cost function during the current iteration, we increase it by a factor of $\xi$. Conversely, if the current learning rate is too large, causing the update step to fail in improving the objective function, we iteratively reduce it by a factor of $\tau$ until a valid update step is found. Finally, if the learning rate shrinks to very small values below a certain threshold, the algorithm is halted. We summarize this procedure as Algorithm~\ref{alg:cap2} in Appendix~\ref{App:alg}.
         
\section{Results}
\label{sec:III}
In this section, we present the results of the periodic orbit search with the tangent space gradient descent algorithm. First, in Sec.~\ref{sec:Ising} we introduce the kicked Ising model used in our example. Next, in Sec.~\ref{sec:res}, we present the periodic orbits identified by our algorithm for different values of bond dimension, analyze their stability, and visualize the KAM tori surrounding the stable orbits. After this, in Sec.~\ref{sec:J} we study the deformation of the periodic orbits as we vary the coupling strengths of the Ising model and discuss the relation between periodic orbits and the eigenstates of the unitary propagator of the quantum model.

\subsection{Kicked Ising model}\label{sec:Ising}
To demonstrate the application of our general algorithm for finding periodic orbits, we must choose a specific quantum model. Specifically, we consider a translationally invariant chain of spin-1/2 degrees of freedom, whose dynamics are described by the kicked Ising model (to avoid confusion, we reserve the use of the term ``Floquet'' to describe properties of periodic orbits except in Sec.~\ref{Sec:discuss}). The model can be described by the following translationally invariant Hamiltonian that is periodic in time: 
\begin{equation}
    H(t) =
    \begin{cases}
        H_1 = \sum_i J \sigma_i^z\sigma_{i+1}^z + h \sigma_i^z,  & \text{$t$ mod $T \in [0,T/2)$}, \\
        H_2 = \sum_i g \sigma_i^x, & \text{$t$ mod $T  \in [T/2, T),$}
    \end{cases}
    \label{Eq:kicked-Ising}
\end{equation}
where $\sigma_i^{x,y,z}$ are Pauli matrices. In our work, we fix the $z$-direction magnetic field $h = 1$, set the period $T=1$, and choose the two remaining couplings to be equal to each other, $g = J$, leaving only a single parameter ($J$). The quantum dynamics of this model over one period is described by the unitary operator
\begin{equation}
    U_T = \mathop{\rm exp}\!\Big(-i\frac{1}{2}H_2\Big)\mathop{\rm exp}\!\Big(-i \frac{1}{2}H_1\Big). 
    \label{eq:Kick}
\end{equation}
This will be used below to compare the eigenvalues of $U_T$ to the periodic orbits identified from TDVP.  

We note that the TDVP algorithm and our orbit-finding method are not sensitive to the time dependence of the Hamiltonian and can be applied in both time-constant and time-dependent settings. However, considering a driven system where Hamiltonian is time-periodic has the advantage of fixing the period of the periodic orbits to be multiples of the driving period. Therefore, the driven setting slightly simplifies the search for periodic orbits, by removing the requirement to optimize for an \emph{a priori} unknown orbit period. An additional advantage of considering the driven system is the absence of any conservation laws, as the time dependence of the Hamiltonian breaks the conservation of energy. As a result, the kicked Ising model in Eq.~(\ref{Eq:kicked-Ising}) has only discrete spatial inversion symmetry, and no other conservation laws or symmetries. 

The kicked Ising model~(\ref{Eq:kicked-Ising}) was actively studied in the literature in the context of quantum thermalization and has several special points in the parameter space. First, when $J = g = \pi/2$ the model~(\ref{Eq:kicked-Ising}) is known to be maximally chaotic. This is the so-called dual unitary point, where the two-site unitary gates that constitute the building blocks of unitary propagator $U_T$ have a special property of maintaining unitarity when one swaps spatial and temporal directions~\cite{akila2016particle, bertini2018exact, bertini2019exact, bertini2019entanglement, piroli2020exact}.  The dual unitary property allows one to obtain exact results for certain properties of the model, such as the correlation functions \cite{bertini2019exact} and spectral form factor \cite{bertini2018exact}. These exact results suggest that, at this point, parts of the system act as an ideal Markovian bath for the remaining degrees of freedom~\cite{lerose2021influence}.         

Upon reducing the coupling $J=g$ away from the unitary point, $J = g = \pi/2$, the model no longer admits treatment with analytic techniques. Nevertheless, numerical studies~\cite{lerose2021influence} suggest that the model stays chaotic for coupling $J=g$ not too far away from $\pi/2$. This is witnessed by the properties of eigenstates of the unitary operator, $U_T$, defined in Eq.~(\ref{eq:Kick}). The eigenstates, encoding infinite time properties of the system, feature large volume-law entanglement and small expectation values of all local observables, consistent with the expectations from thermalization to infinite temperature \cite{d2013many, d2016quantum, lazarides2014equilibrium, mori2016rigorous, mori2018thermalization}, see also Appendix~\ref{App:Preth}. Upon decreasing the coupling further, due to the fact that we keep the period fixed, the model enters the so-called prethermal regime where the coupling $J=g$ takes the role of an effective small parameter (corresponding to high frequency). In this regime, an effective prethermal Hamiltonian can describe the system's behavior over extended periods before heating effects drive the system towards infinite temperature \cite{neyenhuis2017observation, ho2023quantum}. The prethermal phase depends on system size, and at sufficiently large system sizes, it ceases to exist. 

Recently, additional insights into the prethermal regime and the existence of an effective prethermal Hamiltonian were obtained using geometric Floquet theory \cite{schindler2024geometric}.
Ref.~\cite{schindler2024geometric} reveals that many of the long-standing features of periodically driven systems, such as quasienergy folding and ambiguities in state ordering, come from a broken gauge symmetry in the adiabatic gauge potential. Related research shows that the kicked Ising Hamiltonian maintains a low entangled eigenstate -- ground state of the effective Floquet Hamiltonian -- that avoids heating up to some values of $J$~\cite{ikeda2024robust}. In the following, we will argue that our method for finding periodic orbits gives complementary insights into the prethermal regime, and discuss our findings in the context of the earlier works. 

\subsection{Periodic orbits, stability, and KAM tori}
\label{sec:res}

We use the tangent space gradient descent algorithm introduced in Sec.~\ref{sec:GD} to find periodic orbits for the kicked Ising Hamiltonian Eq.~(\ref{eq:Kick}). We choose an intermediate value of the coupling, $J=1.09$ ($J=g$), that puts the model in the prethermal regime (see Appendix \ref{App:Preth}). For each bond dimension $\chi \in [1,4]$ we run the tangent space gradient descent optimization of the fidelity~(Eq.~\eqref{eq:F}) using over one hundred random initial conditions~\footnote{For $\chi=1$ our algorithm essentially reduces to the dynamical mean-field theory, but for consistency we use the same code.}.
For small bond dimensions, $\chi = 1,2$, we use TDVP with a small time step $dt = 0.001$, but for $\chi = 3, 4$ for increased efficiency, we first find approximations for periodic orbits with $dt = 0.01$ and then run a second, refining tangent space gradient descent optimization with a smaller TDVP time step $dt = 0.001$, to converge to the desired precision.

From the converged instances of the tangent space gradient descent optimization, we select instances where the optimized cost function differs from one by less than $10^{-9}$. For small bond dimensions ($\chi = 1,2$), most of the random initializations converge to such values. Notably, with increasing bond dimension, we obtain more instances converging to local minima with fidelity being significantly below one. Hence, the success rate of converging to a periodic orbit decreases with increasing bond dimension. To filter out the duplicate orbits, we analyze the fidelity between all pairs of initial points of converged periodic orbits, ${\cal D}(A_L, A'_L)$ defined in Eq.~(\ref{eq:distance}), and identify duplicate orbits if ${\cal D}(A_L, A'_L)> 1 - 10^{-5}$. Additionally, we discard near-singular orbits, where the second-largest eigenvalue of the transfer matrix ${\rm TM}(A_L,A_L)$ from  Eq.~\eqref{fig:TM} exceeds $1 - 10^{-6}$, as our algorithm assumes injective MPS.

\begin{figure}[t]
    \centering
    \includegraphics[width=\linewidth]{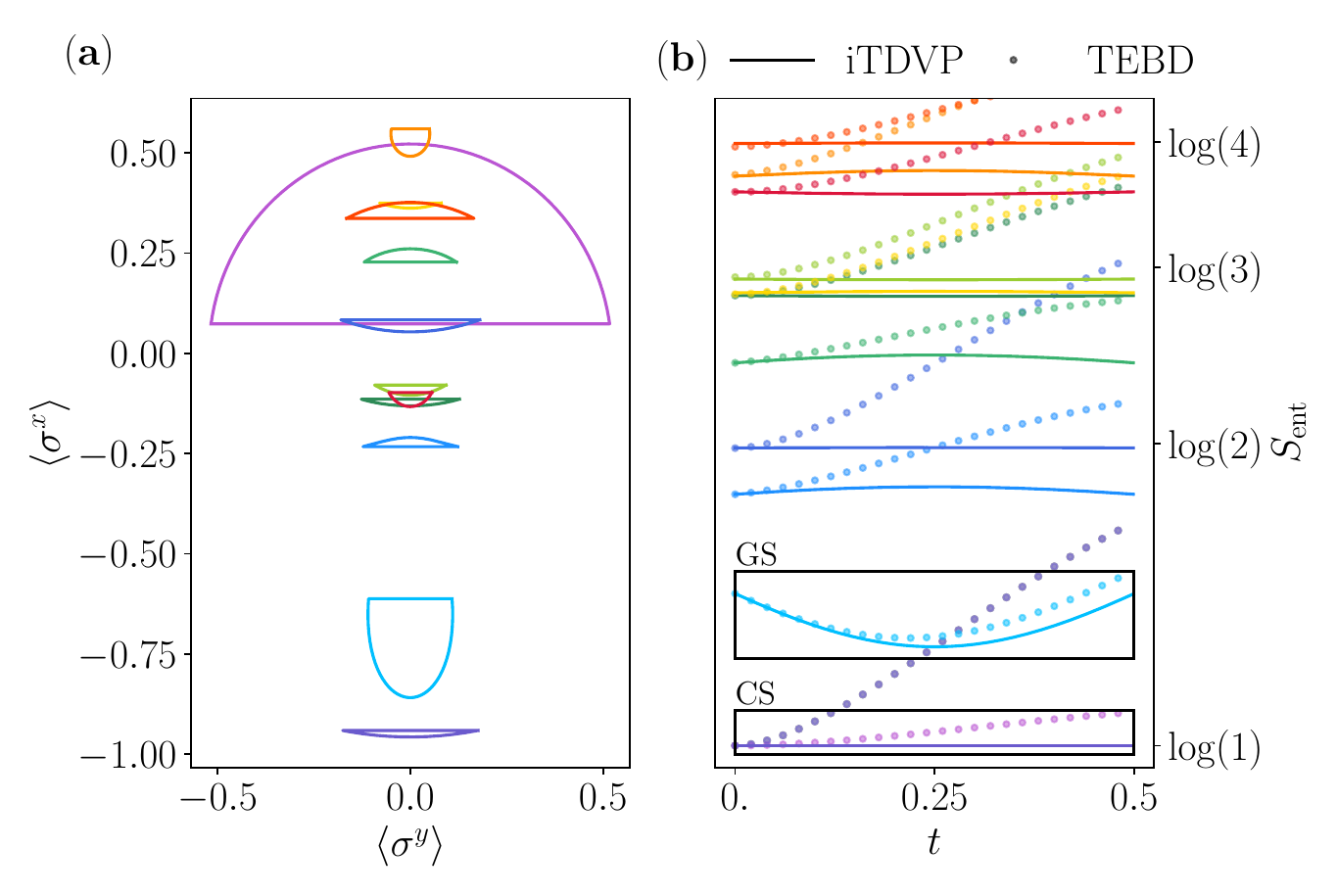}
    \caption{(a) Dynamics of local expectation values $\langle \sigma^{x,y}\rangle$  along periodic orbits of different bond dimensions. Two different orbits for $\chi = 1$ are represented by violet colors, and three unique orbits for $\chi = 2$, and $4$ are shown in different shades of blue and red, respectively, finally 4 orbits for $\chi = 3$ are shown in shades of green. (b) All orbits for $\chi>1$ are characterized by non-vanishing entanglement that is a periodic function of time. Entanglement is shown only during the half-period since it is not affected by the evolution with the single site $x$-magnetic field during the second half-period. Dots correspond to the dynamics of entanglement in full (numerically exact) unitary dynamics obtained with TEBD. All orbits are obtained for $J = 1.09$. The two orbits in the boxes are marked as the ground state (GS) and the ceiling state (CS), the reason for which we discuss in the next section. }
    \label{fig:traj1}
\end{figure}

\begin{figure*}[t]
    \centering
    \includegraphics[width=0.99\linewidth]{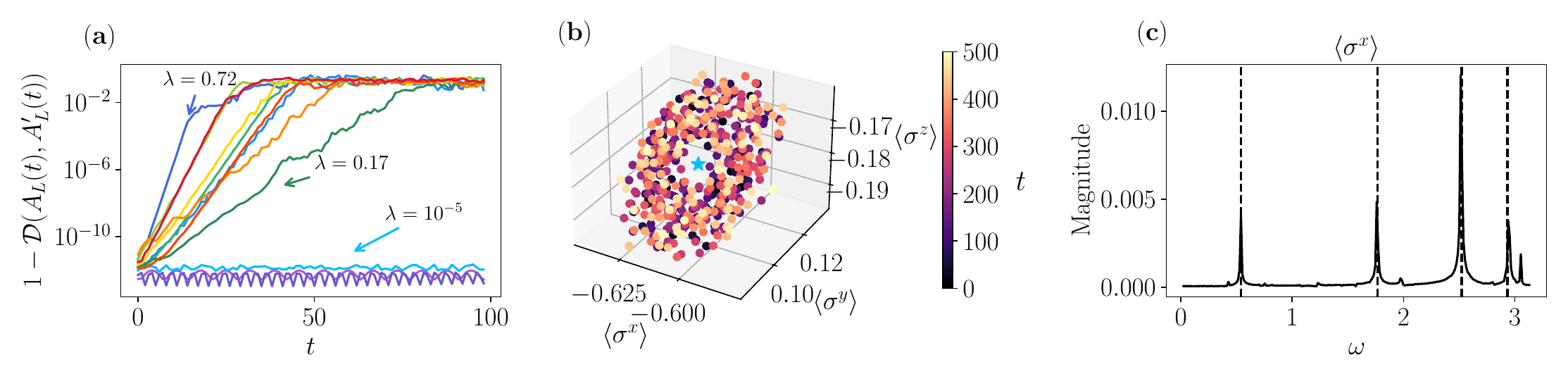}
    \caption{(a) Distance between the time-evolved initial state on the periodic orbit and the trajectory obtained by a small deformation of the orbit, quantified by $ 1 - \mathcal{D}(A_L(t), A_L'(t))$. Trajectories and their color coding coincide with orbits in Fig.~\ref{fig:traj1} for $J = 1.09$. Both $\chi = 1$ orbits and one of the $\chi = 2$ orbits are stable, while the remaining orbits are unstable, and small deformations lead to an exponential increase in distance with time. Floquet exponents for periodic orbits starting from  $\chi = 2$ have the following values $[10^{-5}, 0.3, 0.72, 0.35, 0.17, 0.45, 0.35, 0.29, 0.35, 0.48]$. (b) Stroboscopic dynamics of expectation values of local spin projections for the GS perturbed by a random perturbation with strength $\delta = 0.01$ may be understood as a projection of the KAM torus. The blue star here corresponds to the original periodic orbit values of the local observables. (c) Fourier spectra (without the $\omega = 0$ component) of the $\langle \sigma^x(t) \rangle$ expectation values have sharp peaks that can be interpreted as the frequencies associated with the motion on the four-dimensional KAM torus. Black dashed lines correspond to phases of the four Floquet multipliers defined in Eq.~\eqref{Eq:F-mult} match the location of the peaks (we ignore the small peak at $\omega \approx 3.05$ as it is perturbation-dependent and does not appear for all observables).
    }
    \label{fig:stab}
\end{figure*}

Periodic orbits obtained after the post-processing procedure described above are shown in Fig.~\ref{fig:traj1} for different bond dimensions. 
First, in Fig.~\ref{fig:traj1}(a) we show the expectation values of local observables $\langle\sigma^{x,y}(t)\rangle = \langle \psi(t)|\sigma^{x,y}|\psi(t)\rangle$, where $\psi(t)$ is an orbit at time $t$ in the form of iMPS. The perfect return of the quantum wave function to itself after one period results in closed curves. Moreover, the part of periodic evolution from Eq.~\eqref{Eq:kicked-Ising} where dynamics is generated by the $x$-magnetic field results in straight segments, where $\langle\sigma^x\rangle$ is unchanged. While the presence of entanglement is not apparent in Fig.~\ref{fig:traj1}(a), the next panel (Fig.~\ref{fig:traj1}(b)) illustrates that all orbits with $\chi>1$ have nonzero bipartite entanglement at all times. The relatively poor agreement between the entanglement dynamics on the orbit with the entanglement dynamics from exact simulations (we use TEBD \cite{vidal2003efficient} from the \texttt{ITensor}~\cite{ITensor} package for $L = 50$ and $\texttt{cutoff} = 10^{-32}$) observed for most periodic obits indicates that these orbits are ``high leakage''. In other words, they provide a relatively poor approximation of the exact unitary dynamics. Two orbits, which we call the ground state (GS) and the ceiling state (CS) in Fig.~\ref{fig:traj1} (we discuss the meaning of the names in the next section), have the smallest leakage. The high leakage of the periodic orbits obtained above may be viewed as instability with respect to the degrees of freedom outside the MPS manifold with a fixed bond dimension. We defer further investigation of leakage and its effect on the correspondence between TDVP and full unitary dynamics until Sec.~\ref{sec:J}, and instead turn to the investigation of the stability of the orbits within a fixed-$\chi$ MPS manifold.

To study the stability of periodic orbits with respect to small perturbations in the initial state, we look at the dynamics initiated in the vicinity of the periodic orbit. Specifically, for each orbit, we perturb the initial MPS tensor randomly (by perturbing the individual tensor entries by small random values) with a perturbation of size $\Delta = 10^{-7}$, and visualize the resulting long-time dynamics in Fig.~\ref{fig:stab}. First, we observe that the distance between the state on the periodic orbit and the trajectory resulting from the perturbed orbit has two qualitatively different behaviors, see Fig.~\ref{fig:stab}(a). For most orbits, this distance shows an exponential increase with time, implying that they are unstable and can be associated with a non-zero Lyapunov exponent. However, a few orbits show robustness with respect to random perturbations, where the distance between the orbit and the perturbed state remains bounded with time, suggesting stability. The coexistence of unstable and stable orbits provides evidence for a \emph{mixed phase space}, which was earlier reported only for restricted MPS ansatzes with bond dimension $\chi=2$~\cite{michailidis2020slow}. 

To quantify the (in)stability of periodic orbits, we compute their Floquet multipliers, which measure how a small volume around the initial point of an orbit deforms after one cycle of evolution. Since the TDVP dynamics is symplectic, the phase-space volume is preserved. As a result, the product of all Floquet multipliers is one, reflecting the conservation of phase space volume. Mathematically, the set of Floquet multipliers, $\Lambda$, is defined via the eigenvalues of the Jacobian matrix resulting from the dynamics over the period of the orbit $T$, 
\begin{equation}
    \Lambda = \mathop{\rm eig} J, 
    \  
    J_{ab}={\cal T} \exp\left[\int_0^T d t \frac{\partial {\dot  A}_{L,a}}{\partial A_{L,b}}\right]_{A_L \rightarrow A_L(t)}.
    \label{Eq:F-mult}
\end{equation}
Here we schematically denoted our dynamical variables by $A_{L,a}$ and $A_{L,b}$ (see discussion below for details). For stable orbits, we expect $|\Lambda_i| = 1$, implying that such a box, after one period, rotates but maintains its dimensions. In this case, the eigenvalues come in complex conjugate pairs, and the phases of $\Lambda_i$ from each pair give the frequency of motion along the KAM-torus surrounding the orbit. In contrast, the box surrounding the unstable orbit will expand in the direction corresponding to an eigenvalue outside the unit circle, $|\Lambda_i|>1$, and shrink corresponding to an eigenvalue within the unit circle, $|\Lambda_i|<1$. We also can use the largest-magnitude Floquet multiplier, $\Lambda_\text{max}$, to define an effective Floquet exponent, 
\begin{equation}\label{Eq:F-exp}
    \lambda = \frac{1}{T}\ln|\Lambda_\text{max}|,
\end{equation}
that loosely corresponds to the period-averaged Lyapunov exponent, akin to a finite-time or local Lyapunov exponent \cite{Abarbanel1991,Aurell1997}. In general, we expect that the most unstable Floquet exponent has a dominant influence on the evolution of trajectories near the solution at longer times.

In the present case, where the dynamics is generated by TDVP projection onto the MPS manifold, we use a tangent space basis to remove the unphysical directions corresponding to gauge transformations and construct the Jacobian~(\ref{Eq:F-mult}) numerically (see details in Appendix \ref{App:Jac}). Intuitively, this corresponds to indices $a,b$ in Eq.~(\ref{Eq:F-mult}) running over tangent space directions. Figure~\ref{fig:stab}(a) lists a few values of resulting Floquet exponents $\lambda$, see Eq.~\eqref{Eq:F-exp}, obtained from numerical diagonalization of the Jacobian, illustrating that larger values of $\lambda$ indeed correspond to orbits that are more unstable. 

For stable orbits, in agreement with expectations, our numerical study yields small values $\lambda \approx 10^{-5}$, which are consistent with zero. To further gain insights into the vicinity of such stable periodic orbits, we perturb the stable orbit for $\chi = 2$ and plot the dynamics of local spin excitation values, $\langle\sigma^{x,y,z}\rangle$, at stroboscopic times $nT$, $n \in [1,500]$. Figure~\ref{fig:stab}(b) shows that these stroboscopic expectation values stay in the vicinity of the expectation value corresponding to the periodic orbit for long times. This is consistent with a mixed phase space, commonly observed in Hamiltonian dynamics \cite{ott2002chaos}, where stable orbits are expected to be surrounded by KAM tori~\cite{kolmogorov1954conservation, arnold2009proof, moser1962invariant}. Interpreting $\chi^2$ non-trivial \emph{complex} parameters of the MPS as $\chi^2$ canonically conjugate pairs of \emph{real} momenta and coordinates, we suggest that in the general case, the dimensionality of the torus is equal to $\chi^2$. In Fig.~\ref{fig:stab}(c), we show the Fourier spectra of the observable $\langle \sigma^x(nT)\rangle$. The frequency peaks coincide well with the frequencies obtained from the Floquet multipliers of the stable periodic orbit, meaning that the dynamics of the observables at stroboscopic times $nT$ is consistent with the motion on the surface of a $\chi^2=4$ dimensional KAM torus. In Appendix~\ref{App:stab_9}, we also show evidence of a 9-dimensional torus for stable the $\chi=3$ periodic orbit. 

\begin{figure*}[t]
    \centering
    \includegraphics[width=1.\linewidth]{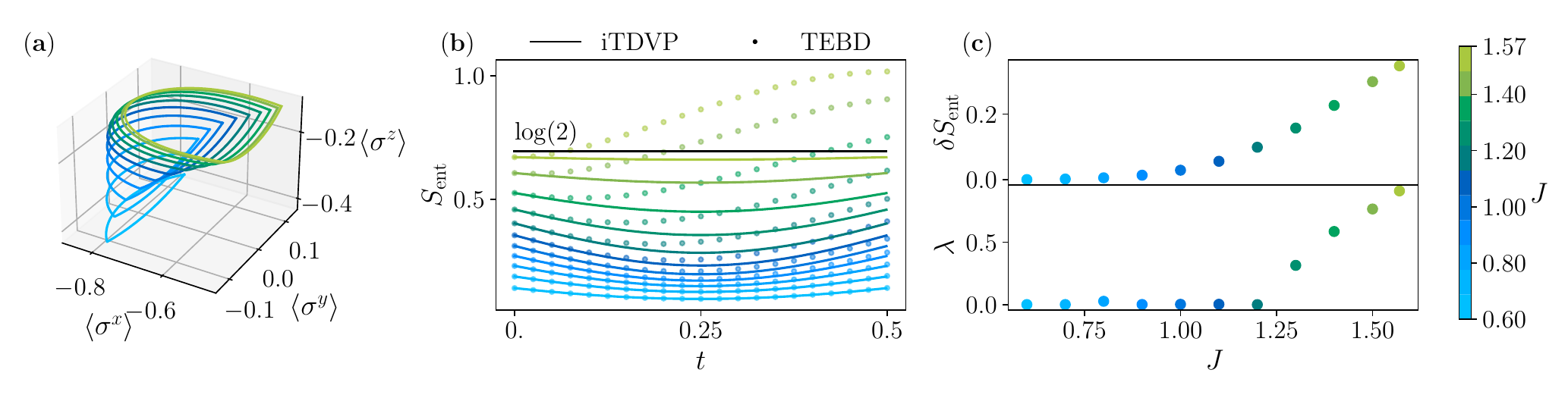}
    \caption{(a) Smooth evolution of the low leakage GS orbit with respect to the propagator parameter $J$, visualized via dynamics of local spin expectation values. (b) Dynamics of entanglement entropy for the first half period of the orbit and from exact unitary dynamics for $L = 50$ TEBD. As we increase $J$, the orbit moves to a more entangled region of the MPS manifold, and the leakage along the orbit increases. (c) The difference between the entanglement entropy in exact unitary and TDVP dynamics ($\delta S_\text{ent}$) at $t = T$ (top panel), and the corresponding Floquet exponent (bottom panel). We observe that both quantities behave in a similar way as functions of $J$. At $J \approx 1.27$, the orbit becomes unstable, and for similar values of $J$, $\delta S_\text{ent}$ starts increasing more rapidly.} 
    \label{fig:J}
\end{figure*}

\subsection{Deformation of orbits and their relation to eigenstates with the Ising coupling}
\label{sec:J}
After analyzing the set of orbits across different bond dimensions for a fixed value of coupling, we consider the fate of the most stable $\chi=2$ orbit when changing the parameter $J = g$. The expectations from classical dynamical systems suggest that the orbit should be smoothly changing with $J$. This is confirmed in Fig.~\ref{fig:J}(a), which shows the smooth change in the dynamics of local spin expectation values with $J \in [0.6, \pi/2]$. Fig~\ref{fig:J}(b) shows the evolution of the entanglement dynamics for the same family of orbits. We observe that, as $J$ changes from a small value to values close to the maximally chaotic dual-unitary point, $J=\pi/2$, the entanglement on the orbit approaches $\ln 2$ -- the maximal value allowed by the fixed bond dimension $\chi=2$. Moreover, the disagreement between TDVP and exact unitary dynamics increases upon increasing $J$.  

Figure~\ref{fig:J}(c) quantifies the leakage of the orbit using the differences in the entanglement entropy dynamics between exact and TDVP evolutions and also shows the evolution of the Floquet exponent, $\lambda$, with the coupling. The orbit becomes unstable at $J\approx 1.27$, where $\lambda$ begins deviating from zero (see Appendix~\ref{App:stab_J} where we discuss the additional insights into this instability from the full analysis of Floquet multipliers in the complex plane). Around somewhat smaller values of the coupling $J$, the leakage also begins to increase more rapidly, as witnessed by the entanglement dynamics. The fact that these phenomena occur \emph{simultaneously} with the entire orbit moving to more entangled regions of the $\chi=2$ MPS manifold highlights the potential relation between the average entanglement on the orbit, its stability, and leakage. Intuitively, we expect that orbits in nearly maximally entangled regions of a given MPS manifold, with $S_\text{ent}$ approaching its strict upper bound, $\ln \chi$, to feature higher leakage and to typically be unstable. Indeed, in this region, the bond dimension is nearly saturated and the MPS ansatz cannot capture additional correlations created by unitary dynamics, giving rise to high leakage.

\begin{figure}[t]
    \centering
    \includegraphics[width=0.98\linewidth]{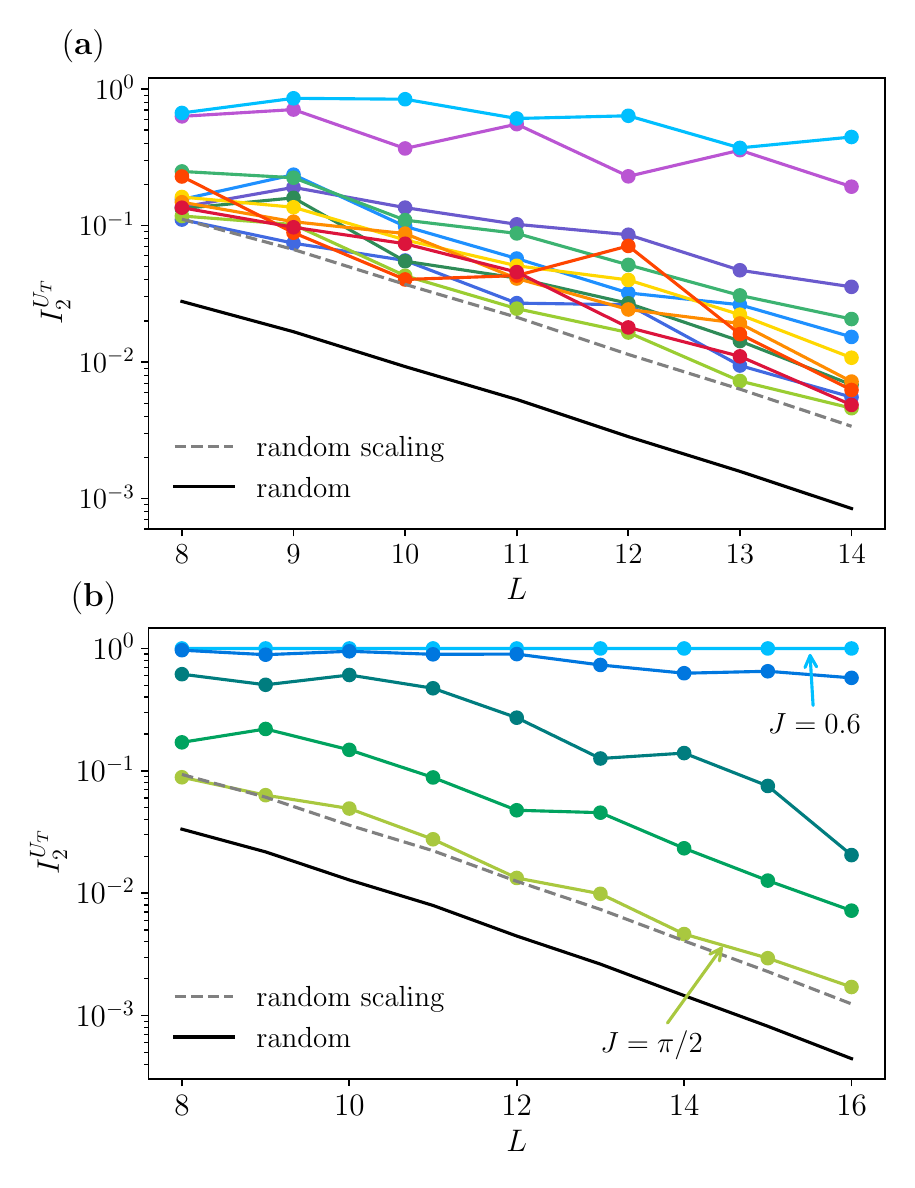}
    \caption{(a) IPR of MPS states corresponding to periodic orbits from Fig.~\ref{fig:traj1} for $J=1.09$ over the basis of eigenstates of the unitary from Eq.~\eqref{eq:Kick}.
    For all cases, IPR decays exponentially with $L$, but two orbits with the lowest leakage show parametrically slower decay of IPR. 
    (b) IPR scaling with $L$ for the lowest leakage $\chi=2$ orbit (GS) at 5 different values of the coupling $J = 0.6, 1.0, 1.2, 1.4, \pi/2$. For $J = 0.6$, the IPR is close to one and stays nearly constant, whereas at the dual unitary point $J=\pi/2$, the IPR decays as $\sim e^{-0.55L}$, in agreement with expectations for a random state, $\sim e^{-0.56L}$. 
    Due to MPS being translation- and inversion-symmetric for this family of orbits, we consider the zero momentum sector with parity $+1$.}
    \label{fig:IPR2}
\end{figure}

Additional support to this conjecture may be obtained by analyzing the effect of the periodic orbit on the exact unitary dynamics of the kicked Ising model generated by the unitary operator in Eq.~\eqref{eq:Kick}. Specifically, we consider the inverse participation ratio (IPR) of the MPS corresponding to the initial point of the orbit,
\begin{equation}
    I_{2}^{U_T} = \sum_{i=1}^D |\langle \psi_\text{MPS}| \varphi_i \rangle|^{4},
\end{equation}
defined over the complete basis of eigenstates $\{| \varphi_i \rangle\}_{i=1}^D$ of the unitary operator, where $D$ is the Hilbert space dimension. The IPR quantifies how delocalized the MPS is on the basis of eigenstates, and in the case where the eigenstates of the unitary operator are fully chaotic, the weakly entangled MPS is expected to show IPR scaling as the inverse size of the Hilbert space dimension, $I_2\sim {1}/{D}\sim e^{-sL}$, thus decreasing exponentially with the system size.

Figure~\ref{fig:IPR2}(a) shows the IPR in the basis of eigenstates for all orbits found for a fixed value of the coupling $J=1.09$, see Fig.~\ref{fig:traj1}. Note that the Hilbert space dimension $D$ corresponds to the size of the translationally invariant zero-momentum sector (we do not fix the parity sector since not all orbits are parity symmetric, see Appendix~\ref{App:inv}). While the majority of MPS states corresponding to periodic orbits show the expected exponential scaling of the IPR with $L$,  MPS states corresponding to low-leakage $\chi=1$ and $2$ orbits in Fig.~\ref{fig:traj1} have IPR close to one that is slowly decreasing with $L$. These two states also appeared to be the ground state (GS) ($\chi = 2$ low-leakage orbit) and ceiling state (CS) ($\chi = 1$ low-leakage orbit) of the effective Hamiltonian, see Appendix~\ref{App:quant}, justifying their names, which were introduced earlier.

Focusing on the GS in Fig.~\ref{fig:IPR2}(b), we show how its IPR scaling changes with the coupling $J$. The evolution of the IPR scaling with $J$ should be compared to the dependence of leakage on $J$ of the same orbit shown in Fig.~\ref{fig:J}(c). At the smallest values of $J$, where the orbit has very small leakage, the IPR is close to one and shows almost no decay with the system size, $L$. This implies that at small values of $J$, our gradient-based orbit search is capable of obtaining approximate eigenstates. From this perspective, our method bears similarity to the density matrix renormalization group (DMRG) approach~\cite{white1992density, verstraete2023density} for finding ground states of local Hamiltonians. In DMRG, the MPS approximation to the ground state is obtained by minimizing the energy of the Hamiltonian, and the quality of the eigenstates can be checked by computing the variance of the Hamiltonian for the approximate MPS eigenstate. Similarly, in our case, we maximize the fidelity~(\ref{eq:F}) to find periodic orbits and check their quality via the leakage~(\ref{eq:leak}). As a result, the low-leakage orbits can be interpreted as approximate eigenstates of the unitary operator generating dynamics over one period, with the dynamics on the periodic orbit corresponding to the micromotion (dynamics happening inside the period) of the driven system with the deviation bounded by the leakage~(\ref{eq:leak}).

From the discussion above, we conclude that low-leakage periodic orbits imply the existence of weakly entangled eigenstates of the unitary operator~(\ref{eq:Kick}). General many-body systems are not expected to have such low-entangled eigenstates. Rather on the contrary, all eigenstates are expected to realize an infinite temperature ensemble \cite{d2013many, d2016quantum, lazarides2014equilibrium, mori2016rigorous, mori2018thermalization}.  However, as discussed in Section~\ref{sec:Ising}, for small coupling $J$, the kicked Ising model features a prethermal regime. Here, the dynamics shows long-lived, quasi-stationary behavior before eventually thermalizing. The system effectively evolves as if governed by an approximately conserved effective Hamiltonian \cite{neyenhuis2017observation, ho2023quantum}, leading to the existence of eigenstates with small entanglement. Thus, prethermal behavior explains the success in finding low leakage periodic orbits for small $J$ (see also Appendix~\ref{App:quant} for additional details).

\section{Discussion}\label{Sec:discuss}
In this paper, we introduced a new method for finding periodic orbits in the TDVP equations of motion projected onto the manifold of MPS for general bond dimensions. Our method maximizes the fidelity using gradient descent, therefore having complexity parametrically larger compared to TDVP, $O(d\chi^5)$ vs $O(d\chi^3)$, with the additional factor emerging due to the computation of $\chi^2$ gradients. Applying this method to the kicked Ising model, we found a number of periodic orbits for a range of bond dimensions $\chi \in [1,4]$. We demonstrated that various techniques from classical chaos can be also applied to the variational projection of quantum dynamics. In particular, we performed orbit stability analysis via Floquet multipliers, visualized the KAM tori surrounding the stable orbits, and characterized the dimension of KAM tori via Fourier transforms of time series of local observables. 

By adjusting the Ising coupling, we tuned the quantum dynamics between prethermal and chaotic regimes. At the same time, we tracked the fate of periodic orbits and observed their smooth deformation, and the onset of instability upon approaching the maximally chaotic point. In the prethermal regime, our method for finding orbits results in approximate eigenstates of the unitary operator generating the exact dynamics over one period and, thus, may be viewed as a generalization of the DRMG to the periodically driven setting. In the chaotic regime, we observe that although periodic orbits survive, they become unstable and no longer correspond to approximate eigenstates. It remains to be seen if they still leave imprints on quantum dynamics, despite their high leakage. 

TDVP projection of quantum dynamics onto the MPS manifold was known to result in a classical dynamical system with a symplectic structure. Yet, the resulting classical dynamics have not been studied in detail until now, with the exception of a few works that analyzed the spectrum of Lyapunov exponents resulting from long-time dynamics~\cite{hallam2019lyapunov} and analytically studied TDVP projection for a special class of MPS states with a small number of parameters~\cite{ho2019periodic,michailidis2020slow}. The main achievement of the present work is the demonstration that such dynamics, despite the complex nature of dynamical variables, can be systematically studied with tools from classical chaos. The key insight, consistent with the general approach to MPS \cite{vanderstraeten2019tangent}, is to consider only gauge invariant quantities, such as fidelity and local observables, and use these to infer the properties of the classical dynamical system, while never unpacking the dynamical variables, leaving them hidden inside MPS tensors.

The possibility of obtaining the classical limit of quantum dynamics with TDVP projection onto the MPS manifold opens a number of exciting directions. First, a practical application of our algorithm is the systematic search for non-thermal eigenstates that may appear at high energy densities in Hamiltonian systems and are known as quantum many-body scars~\cite{Turner2017,Serbyn:2021vc,Moudgalya_2022,chandran23,HummelPRL,Dag24PRL,ermakov2024,pizzi2024,Dag24,LjubotinaPRXQ}. The appearance of scars was linked to periodic orbits with sufficiently slow entanglement growth (small leakage)~\cite{ho2019periodic, michailidis2020slow}. Critically, our algorithm opens the door to the search of such orbits within MPS manifolds of general bond dimension. Moreover, by finding such periodic orbits in regions characterized by small leakage, it may be possible to observe quantum signatures of phenomena established in classical dynamical systems, such as bifurcation transitions in the periodic orbits, Arnold diffusion~\cite{vi1974instability,Lichtenberg1983}, and others.

Dynamics of other classes of quantum systems that avoid thermalization is another attractive avenue for the application of our method. In particular, a straightforward extension of our algorithm to finite-size MPS simulations without translational invariance will open the door to the study of the kicked Ising model with disorder -- one of the models realizing Floquet many-body localization~\cite{zhang2016floquet, ponte2015many,lazarides2015fate,d2013many, abanin2016theory}. It would be interesting to check if the present algorithm is capable of finding eigenstates within the Floquet many-body localized phase for system sizes beyond those that can be reached with exact diagonalization. It also remains to be understood, if the classical phase portraits in the many-body localized phase feature large regular regions of the phase space, thereby directly establishing an analogy between many-body localization and the Kolmogorov-Arnold-Moser theorem~\cite{AbaninRMP}. In a different direction, the study of similar questions for Bethe-ansatz integrable models~\cite{sutherland2004beautiful} with translational invariance is also an attractive avenue. For such models, however, additional research is needed as it may need the nontrivial extension of the TDVP that is potentially capable of preserving additional conservation laws. 

Finally, we expect that our approach may also provide useful insights for chaotic, periodically driven, or Hamiltonian systems. In particular, we conjecture that in a chaotic regime of quantum dynamics, periodic orbits are hidden in the regions with nearly maximal entanglement allowed by the MPS manifold and are generally unstable. It remains to be understood if the properties of such orbits can nevertheless provide useful insights into quantum thermalization. 

\emph{Note added.} During the completion of this work, we became aware of a related study by Ren \emph{et al.}~\cite{zlatko}, where a method based on an iterative sequence of finite time evolution and projection back onto an MPS manifold was developed to identify periodic orbits related to quantum many-body scars in PXP and other models.
\begin{acknowledgments}
We acknowledge useful discussions with C.~Kollath, A.~Green, and D.~Huse. 
E.~P., M.~L., and M.~S. acknowledge support by the European Research Council under the European Union's Horizon 2020 research and innovation program (Grant Agreement No.~850899). M.~L. acknowledges support by the Deutsche Forschungsgemeinschaft (DFG, German Research Foundation) under Germany’s Excellence Strategy – EXC-2111 – 390814868. This research was supported in part by grant NSF PHY-2309135 to the Kavli Institute for Theoretical Physics (KITP).
\end{acknowledgments}

\appendix
\section{Implementation of TDVP and gradient descent}
\label{App:1}
In this Appendix, we discuss the construction of the mixed gauge of iMPS, the iMPS tangent bases, and the TDVP algorithm that we implement in our work based on the papers \cite{vanderstraeten2019tangent, zauner2018variational}. After this, we provide the details for the implementation of the gradient descent algorithm introduced in Sec.~\ref{sec:GD} in the main text for the search of closed orbits.

\subsection{Mixed canonical form  \label{App:canonical}}

The first step in numerical algorithms dealing with iMPS is to obtain the mixed canonical form, which fixes part of the gauge freedom, as shown in Eq.~\eqref{fig:LCR}. The algorithm implemented in our code follows the approach described in Ref.~\cite{vanderstraeten2019tangent}, which we briefly review here for completeness.

The concept of gauge fixing arises from the observation that, given an MPS representation as in Eq.~\eqref{fig:AAA}, the tensor $A$ can be transformed via an invertible matrix $X$ according to the rule $A'^s_{ij} = X^{-1}_{im} A^s_{mn} X_{nj}$ (throughout the following, we denote such a transformation compactly as $X^{-1}AX$ unless otherwise specified). Importantly, this transformation leaves the physical state invariant. 
This gauge freedom allows us to impose additional structure on the MPS. In particular, it enables us to choose a transformation $X$ that brings the tensors into a form satisfying either the left or right orthonormalization condition, 
\begin{equation}
\begin{aligned}
  &\text{\raisebox{-10mm}{\includegraphics[width=0.85\linewidth]{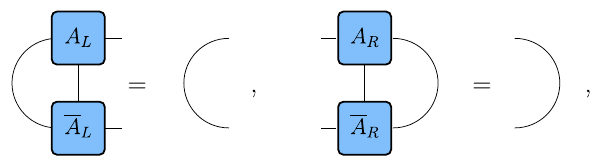}}}
\end{aligned}
\label{Eq:LR}
\end{equation}
therefore partially fixing the gauge.

In this Section, we use the following notation. To find the left canonical form, we call such a transformation $L$, and so have $A_L = LAL^{-1}$, and analogously for the right canonical form, we call the transformation $R$. The matrix $L$ can be found from the left fixed point of the transfer matrix $\text{TM}(A,A)$:
\begin{equation}            
\begin{aligned}
  &\text{\raisebox{-10mm}{\includegraphics[width=0.4\linewidth]{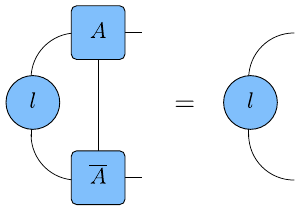}}}
\end{aligned}
\label{eq:Figl}
\end{equation}
by solving the equation $l = L^{\dagger}L$ for $L$. Then the left gauge fixing condition is satisfied as: 
\begin{equation} 
\begin{aligned}
  &\text{\raisebox{-10mm}{\includegraphics[width=0.75\linewidth]{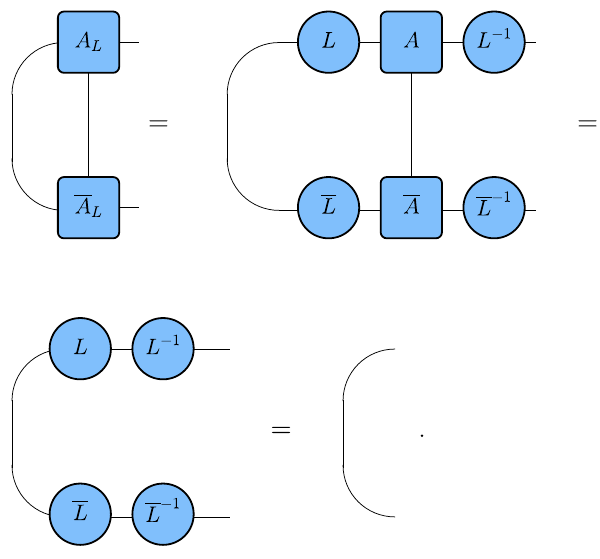}}}
\end{aligned}
\label{eq:Fig_AL}
\end{equation}
The same procedure can be applied to the right canonical form. After finding $A_L$ and $A_R$, the only degrees of freedom that are left are unitary transformations
\begin{equation}
\begin{gathered}
    A'_L = U^{\dagger}A_LU\\
    A'_R = V^{\dagger}A_R V
\end{gathered}
\label{eq:gauge}
\end{equation}
which leave tensors left/right orthonormal. 

The procedure for obtaining the left and right canonical forms of an MPS tensor can be formulated as an iterative process. Given a tensor $A$, our goal is to find a transformation matrix $L$ such that the transformed tensor $A_L = L A L^{-1}$ satisfies the left gauge fixing condition Eq.~\eqref{eq:Fig_AL}. The optimal way to find $A_L$ is to solve $LA = A_L L$.
The iterative algorithm for that proceeds as follows. Starting from an initial guess $L^i$, we compute the product $L^i A$ and perform a QR decomposition on the result. This yields a new tensor $A_L^{i+1}$ and an updated matrix $L^{i+1}$ as 
\begin{equation}
\begin{aligned}
  &\text{\raisebox{-10mm}{\includegraphics[width=0.55\linewidth]{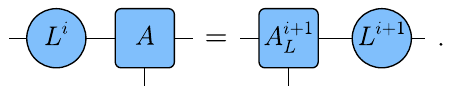}}}
\end{aligned}
\label{eq:Fig_it}
\end{equation}
The updated matrix is then used as the input for the next iteration, and the process is repeated until convergence.

The QR decomposition is unique up to the signs inside the decomposition. That is why we fix the diagonal elements of the triangular matrix to be positive.
The algorithm reaches a fixed point when $L^{(i+1)} = L^{(i)} = L $, and the resulting tensor $ A_L $ is left-orthonormal by construction.
To accelerate convergence, once an intermediate $A_L^{i+1} $ has been found, we improve the next guess for $ L $ by replacing it with a refined fixed point $ \tilde{L}^{i+1} $ of the map:
\begin{equation}
\begin{aligned}
  &\text{\raisebox{-10mm}{\includegraphics[width=0.5\linewidth]{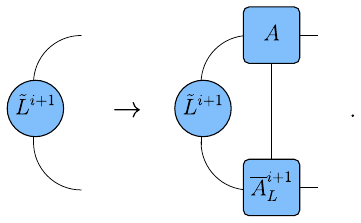}}}
\end{aligned}
\label{eq:Fig_X}
\end{equation}
as the following is satisfied for the left canonical form:
\begin{equation}
\begin{aligned}
  &\text{\raisebox{-10mm}{\includegraphics[width=0.8\linewidth]{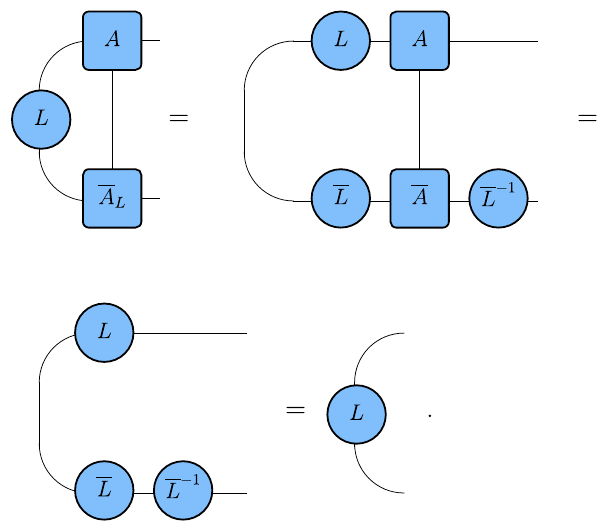}}}
\end{aligned}
\label{eq:Fig_XX}
\end{equation}
The general procedure described above is summarized in  Algorithm~\ref{alg:MC}. For the \texttt{eigensolver}, we used the \texttt{KrylovKit} package in Julia and the \texttt{LAPACK} library in C++, with a convergence tolerance set to $10^{-15}$.
\begin{algorithm}[t]
\caption{Left orthonormal form}
\label{alg:MC}
\algorule
\begin{algorithmic}[1]
\Require {$A, L, \eta$} 
    \State $L = L / \|L\|$ 
    \State $L_{\text{old}} = L$
    \State $(A_L, L) = \texttt{QRGauged}(LA)$ \Comment{Gauged QR decomposition Eq.\eqref{eq:Fig_it}}
    \State $L = L / \|L\|$ 
    \State $\delta = \|L - L_{\text{old}}\|$ 
    \While{$\delta > \eta$} 
        \State $L = \texttt{eigensolver}(X \rightarrow \text{TM}(X), L, \delta / 10)$ \Comment{Finding fixed point of transfer map in Eq.\eqref{eq:Fig_X}}
        \State $L = \texttt{QRGauged}(L)$ 
        \State $L = L / \|L\|$
        \State $L_{\text{old}} = L$
        \State $(A_L, L) \gets \texttt{QRGauged}(LA)$ \Comment{Gauged QR decomposition Eq.\eqref{eq:Fig_it}}
        \State $L = L / \|L\|$ 
        \State $\delta = \|L - L_{\text{old}}\|$
    \EndWhile
    \State \Return $A_L, L$
\end{algorithmic}
\algorule
\end{algorithm}

The procedure for obtaining the right canonical form mirrors that of the left one. However, instead of starting with the original tensor $A$, we begin with the left-canonical tensor $A_L$. In this case, we perform an RQ decomposition to extract the right canonical form. The central cite $C$ is obtained from $R$, which we have after right orthonormalization by doing gauged SVD on it and transforming $A_L$ and $A_R$ through $U$ and $V$ matrices as in Eq.~\eqref{eq:gauge}. We also do gauge fixing in SVD, as the phases of singular vectors have a phase freedom.  For that, we scan through left singular vectors and make the first non-zero component real and positive, adjusting the phase of the right singular vectors accordingly. The entire procedure for obtaining the mixed canonical form is summarized in Algorithm~\ref{alg:MC2}.

\begin{algorithm}
\caption{Find mixed gauge \(\{A_L, A_R, C\}\)}
\label{alg:MC2}
\algorule
\begin{algorithmic}[1]
\Require {$A, \eta$} 
\State \((A_L, \sim) = \texttt{LeftOrthonormalize}(A, L_0, \eta)\) 
\State \((A_R, C) =\texttt{RightOrthonormalize}(A_L, C_0, \eta)\) 
\State \((U, C, V) = \texttt{SVDGauged}(C)\) 
\State \(A_L = U^\dagger A_L U\) 
\State \(A_R = V^\dagger A_R V\) 
\State \Return \(A_L, A_R, C\)
\end{algorithmic}
\algorule
\end{algorithm}

We notice that the iterative procedure described above may encounter an issue when initialized with an MPS that has a poorly conditioned entanglement spectrum. Intuitively, this corresponds to an initialization with a state of a nominal bond dimension $\chi$, which is nearly identical to the state with a smaller bond dimension $\chi'<\chi$. For such initializations, the procedure of computing $A_R$ was stuck in the infinite loop as $R$ does not converge. Intuitively, the presence of very small singular values in the entanglement spectrum leads to the matrix $R$ having small values on the diagonal, which does not allow fixing the gauge correctly in QR decomposition. This could be partially fixed by restarting the algorithm, using a different initial guess of $R$. However, in cases when several such restarts did not help, we terminate the entire procedure of searching for the closed orbit and use a different initial guess, as the orbit in question can most likely be identified using a smaller bond dimension.

\subsection{Tangent basis and TDVP realization}
\label{App:TDVP}
We start with the construction of the tangent basis for the iMPS in the mixed canonical gauge.
The tangent space for iMPS is defined as the set of all possible infinitesimal changes of the iMPS that correspond to physical changes of the quantum state. 
The most general form of the tangent vector has the following form:
\begin{equation}
\begin{aligned}
  &\text{\raisebox{-1mm}{\includegraphics[width=0.9\linewidth]{Fig1.pdf}}}
\end{aligned}
\label{fig:LVR}
\end{equation}
where tensor $V$ parametrizes directions in the tangent space.
In this form, tangent vectors also have a gauge freedom. Changing $V \to V + A_LB - BA_R$ for any matrix $B$ leaves the tangent vector the same. We could fix this freedom by implementing left gauge fixing:
\begin{equation}
\includegraphics[width=0.3\linewidth]{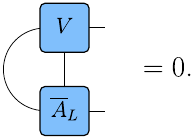}
\label{fig:Bb}
\end{equation}
Following the equation above, we construct the tensor $V$ using a null space of the tensor $A_L$ with merged left bond and physical indices, denoted as $\nu$. $\nu$ is a $d\chi \times (d-1)\chi$ matrix, where a number of columns correspond to the size of the null space of the tensor $A_L$:
\begin{equation}
\centering
\includegraphics[width=0.85\linewidth]{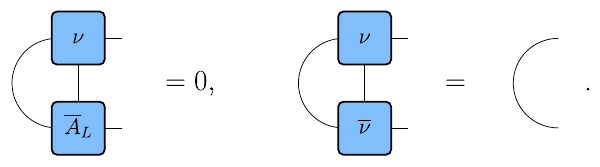}
\label{fig:VL1}
\end{equation}
Having matrix $\nu$ we define unique tangent space directions using a set of $(d-1)\chi \times \chi$ dimensional tensors $\{X^{(\alpha)}\}$ as follows: 
\begin{equation}                        \includegraphics[width=0.55\linewidth]{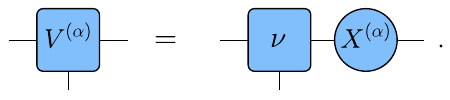}
\label{fig:V}
\end{equation}       
To construct an orthogonal tangent basis, we choose $\{X^{(\alpha)}\}$ as the basis in the space of complex matrices of size $(d-1)\chi \times \chi$. Each matrix $X^{(\alpha)}_{\text{re(im)}}$ in the set has exactly one nonzero element $1$($i$). This choice provides $2(d-1)\chi^2$ independent basis vectors, where the factor of 2 arises from treating real and imaginary parts as separate components. 
Eq.~\eqref{fig:V} defines tangent basis as two sets of tensors $\{V^{(\alpha)}_{\text{re}}\}$ and $\{V^{(\alpha)}_{\text{im}}\}$, which satisfy the relation:
\begin{equation}
i V^{(\alpha)}_{\text{re}}  =  V^{(\alpha)}_{\text{im}} ,
\label{eq:basis_app}
\end{equation}
where index $\alpha$ runs over $\alpha \in [1, (d-1)\chi^2]$.

The tangent space of the iMPS plays the main role in the time-dependent variational principle realization.
The TDVP equation of motion is a projected Schrodinger equation:
\begin{equation}
\partial_t |\psi(A_L, A_C, A_R)\rangle = -i\mathcal{P}H|\psi(A_L,A_C, A_R)\rangle
\label{eq:sch}
\end{equation}
The projector in the TDVP equation maps the Hamiltonian's action back onto the tangent space spanned by the basis in Eq.~\eqref{eq:basis_app}. The most optimal way to project the dynamics on the manifold is to minimize the leakage or the distance between the projected vector and the derivative vector, and the exact form of such a projector could be found in \cite{vanderstraeten2019tangent}. Due to its structure,  the Eq.~\eqref{eq:sch} splits into two equations, one is for tensor $A_C$, and another is for $C$:
\begin{equation}
\begin{gathered}
\dot{A}_C = -iG_1(A_C),\\
\dot{C} = iG_2(C).
\label{eq:ev}
\end{gathered}
\end{equation}
Here, $G_1$ and $G_2$ are linear hermitian maps that take the tensor $A_C/C$ and return a tensor with the same dimensions. They represent the action of the projected Hamiltonian on the MPS. 
The solution of Eq.~\eqref{eq:ev} can be schematically represented as an exponential of these maps $A_C(t) = e^{-iG_1t} A_C(0)$ and $C(t) = e^{iG_1t} C(0)$.
The equations below show the action of $G_1$ and $G_2$ on tensors $A_C$ and $C$:
\begin{equation}
\includegraphics[width=1.\linewidth]{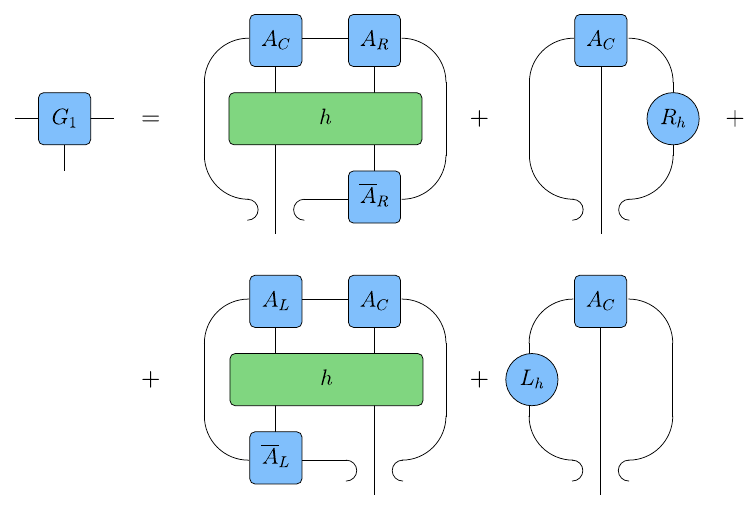}
\label{fig:TDVP1}
\end{equation}
\begin{equation}
\includegraphics[width=0.4\linewidth]{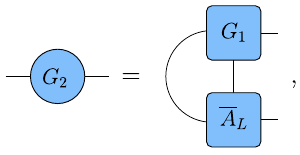}
\label{fig:TDVP2}
\end{equation}
where in Eq.~\eqref{fig:TDVP2} dependence on $C$ is coming from the relation $(A_L)_{ij}^sC_{jk} = (A_C)^s_{ik}$.

In Eqs.~\eqref{fig:TDVP1} and \eqref{fig:TDVP2}, $(L_h|$ and $|R_h$) are the environments on the left and on the right, which represent the case when the Hamiltonian acts to the left side from the updating site and to the right side from the updating site. Eq.~\eqref{fig:Lh} shows the left environment tensor construction:
\begin{equation}
\includegraphics[width=0.99\linewidth]{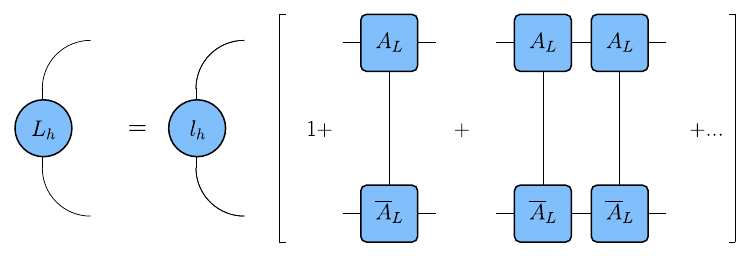}
\label{fig:Lh}
\end{equation}
where $l_h$ is defined as:
\begin{equation}
\includegraphics[width=0.6\linewidth]{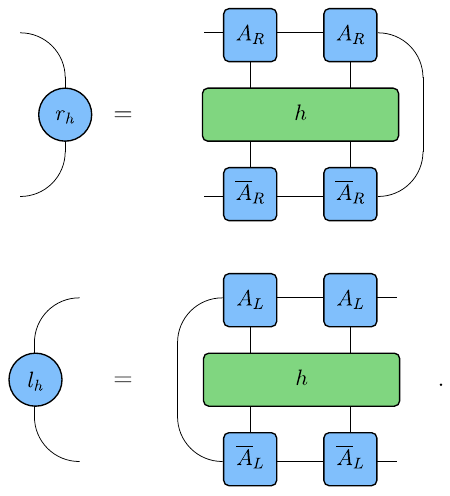}
\label{fig:lh}
\end{equation}
The series in Eq.~\eqref{fig:Lh} and in the analogous equation for $R_h$, could be rewritten in the following form:
\begin{equation}
\begin{gathered}
(L_h| = (l_h| \sum_0^{\infty} \text{TM}(A_L, A_L)^k, \\ 
|R_h) =\sum_0^{\infty} \text{TM}(A_R, A_R)^k|r_h),
\label{eq:Lh}
\end{gathered}
\end{equation}
where the index of the transfer matrix $L/R$ corresponds to the tensors $A_L/A_R$ on which the transfer matrix~\eqref{fig:TM} is constructed.

To find the left and right environments, one needs to sum an infinite number of transfer matrices $\text{TM}_{L/R}$ as in Eq.\eqref{fig:Lh}.
Let us now concentrate on the $(L_h|$ vector for simplicity.
The transfer matrix $\text{TM}(A_L, A_L)$ has a dominant eigenvalue of unit magnitude, with corresponding left and right eigenvectors denoted by $( \mathbb{1}|$ and $|R)$, respectively. The projection $(l_h | \text{TM}(A_L, A_L^k |R ) = ( l_h | R )$ represents the energy density expectation value that doesn't depend on the position of Hamiltonian $k$.
The same logic applies to $| R_h )$.
Then, to find the right and left environments~.\eqref{eq:Lh} we solve two equations:
\begin{equation}
\begin{gathered}
(L_h |  \big[{1} + | R)( \mathrm{1} | - \text{TM}(A_L, A_L) \big] = (l_h | - (l_h | R ) ( \mathbb{1} |,  \\
\big[\mathbb{1}  + |\mathbb{1} )( L |- \text{TM}(A_R, A_R)\big] | R_h ) = | r_h ) - |\mathbb{1})( L | r_h ) 
\end{gathered} 
\end{equation}

At this point, we have all the building blocks to perform time evolution by one time step. According to Eq.~\eqref{eq:ev}, $A_C(\delta t) = e^{-iG_1 \delta t}A_C(0)$ and $C(\delta t) = e^{iG_2 \delta t}C(0)$.
After applying time evolution for time $\delta t$ and obtaining tensors $A_C(\delta t)$ and $C(\delta t)$, we have to extract $A_L(\delta t)$ and $A_R(\delta t)$. For this, we join two indices of $A_C$ to the right (upper index $r$) or the left (upper index $l$) and apply a polar decomposition:
\begin{eqnarray}
A_C^{l} &= U_{A_C}^{l} P_{A_C}^{l}, \qquad
&A_C^{r} = P_{A_C}^{r} U_{A_C}^{r},\\
C &= U_C^{l} P_C^{l},  \qquad
&C = P_C^{r} U_C^{r}. 
\end{eqnarray}
Then the tensors $A_L$ and $A_R$ are obtained as follows:
\begin{equation}
A_L = U_{A_C}^{l} U_C^{l \dagger}, \qquad
A_R = U_C^{l \dagger} U_{A_C}^{r}.
\label{eq:ALR}
\end{equation}

Above, we discussed how to evolve the MPS tensors $A_{L/C/R}/C$ of the system by one time step. Let us now move to the integration procedure for the entire MPS.
To better understand the integration of the TDVP equation for infinite MPS, let us first look at the general finite system size MPS.
To propagate the finite system, the following steps are needed. Consider all $A_L(n-1)$ to the left of $n$ as already evolved by one step, and $A_R(n+1)$ and further not. For the system without translation-invariant property, all tensors are different, and so the tangent space, projector, and maps $G_{1,2}$ are likewise different. The procedure is the following:
\begin{itemize}
\item Evolve central tensor according to the Eq.~\eqref{eq:ev}: $\tilde{A}_C(n) = e^{-iG_1(n)\delta t}A_C(n)$.
\item Do QR decomposition to extract $A_L$ and $C$: $\tilde{A}_C(n) = \tilde{A}_L(n) \tilde{C}(n) $.
\item Evolve tensor $C$ according to Eq.~\eqref{eq:ev}: $\hat{C}(n) = e^{iG_2(n) \delta t}\tilde{C}(n)$.
\item Absorb new $\hat{C}$ to $A_C(n+1) = \hat{C}(n)A_R(n+1)$ and continue to the next site.
\end{itemize}

For an infinite system, integration works somewhat differently. Compared to the procedure above the iMPS should remain translationally invariant for all times $t$. This implies that we should apply the procedure above until $A_C(n) = A_C(n+1)$. Another way to look at this relation is to notice that after the propagation for one time step, we obtain the same matrix $C$ with which we started, $C(t + \delta t) = C(t)$. This we can be used to find $\tilde{C}$ as a \emph{backward evolution} of $C$, because $\hat{C} = C = e^{iG_2 \delta t}\tilde{C}$, then we have: $\tilde{C} = e^{-iG_2 \delta t}C$.
And finally from $\tilde{A}_C$ and $\tilde{C}$ we can obtain $\tilde{A}_L$ and $\tilde{A}_R$.
Thus the time evolution procedure for iMPS can be summarized by the following steps:
\begin{itemize}
\item Time evolve the center-cite tensor forward in time $\tilde{A}_C = e^{-iG_1\delta t}A_C$
\item Time evolve the center-cite tensor backward in time $\tilde{C} = e^{-iG_2\delta t}C$
\item Find and update $\tilde{A}_L$, $\tilde{A}_R$ from $\tilde{A}_C$ and $\tilde{C}$ according to Eq.~\eqref{eq:ALR}.
\end{itemize}

\subsection{Gradient descent algorithm and convergence properties}
\label{App:alg}

\begin{algorithm}[t]
\caption{Iterator for gradient descent optimization}
\label{alg:cap}
\algorule
\begin{algorithmic}[1]
\Require $parameters$, $c$
\State Construct $A$ from $parameters$ 
\State $(A_L,\sim, \sim) = \texttt{MixedCanonical}(A)$
\State $(A_{Lnew}, c) = \texttt{GradientDescent}(A_L, c)$          
\While{$(F(A_{Lnew}) - F(A_L)) > tol$ || $(1 - F(A_{Lnew})) > atol$}\Comment{$F$ from Eq.\eqref{eq:F}}
   \State $A_L = A_{Lnew}$
   \State $(A_{Lnew}, c) = \texttt{GradientDescent}(A_L, c)$
   \State $c = \min(c, max_c)$
\EndWhile
\State \Return $A_{Lnew}$
\end{algorithmic}
\algorule
\end{algorithm}

\begin{algorithm}[b]
\caption{Gradient Descent}\label{alg:cap2}
\algorule
\begin{algorithmic}[1]
\Require $A_L$, $c$
\State $g_{\text{re}/\text{im}}^{\alpha} = \frac{F(A_L + \Delta V_{\text{re}/\text{im}}^{(\alpha)}) - F(A_L - \Delta V_{\text{re}/\text{im}}^{(\alpha)})}{2\Delta}$

\State $\delta A_L = \sum_{\alpha} (g_{\text{re}}^{\alpha} + ig_{\text{im}}^{\alpha})V^{(\alpha)}_{\text{re}}$
\State Normalize $\delta A_L$ \Comment{Using $ \mathcal{D}(A,A)$ from Eq.~\eqref{eq:distance}}
\State $A_{Lnew} = A_L + c \delta A_L$
\State $f = F(A_{Lnew})$ \Comment{$F$ from Eq.~\eqref{eq:F}}
\If{$f > F(A_L)$}
   \State $c = c*\xi$
\EndIf
\While{$f \leq F(A_L)$}
   \State $c = c/\tau$
   \State $A_{Lnew} = A_L + c \delta A_L$
   \State $f = F(A_{Lnew})$
   \If{$c < \epsilon$}
       \State break
   \EndIf
\EndWhile
\State Normalize $A_{Lnew}$  \Comment{Using  $ \mathcal{D}(A,A)$ from Eq.~\eqref{eq:distance}}
\State \Return{$A_{Lnew}$, $c$}
\end{algorithmic}
\algorule
\end{algorithm}

To run the algorithm, we initialize a random MPS by generating a vector of $2d\chi^2$ random numbers in the range $[-1,1]$ and reshaping it to a complex matrix $A$ of size $\chi \times d \times \chi$, which we use to start the optimization. First, we construct the left canonical form and then proceed with the gradient descent optimization of the cost function. The conceptual description of the iterator of the algorithm and initial preparation are combined in Algorithm~\ref{alg:cap}. 

This procedure relies on a tangent space gradient descent subroutine, which is briefly described in the main text Sec.~\ref{sec:GD}. Here we provide a more detailed description of this procedure, summarized in Algorithm~\ref{alg:cap2}.
First, we compute gradients using a tangent space basis. 
These gradients are then used to construct the update tensor $\delta A_L$, which determines the direction of the update. To ensure effective optimization, the learning rate, denoted by $c$, is adjusted as described in Sec.~\ref{sec:GD}, and this adjusted rate is applied to update the tensor $A_L$. 

\begin{figure}[t]
\centering
\includegraphics[width=0.99\linewidth]{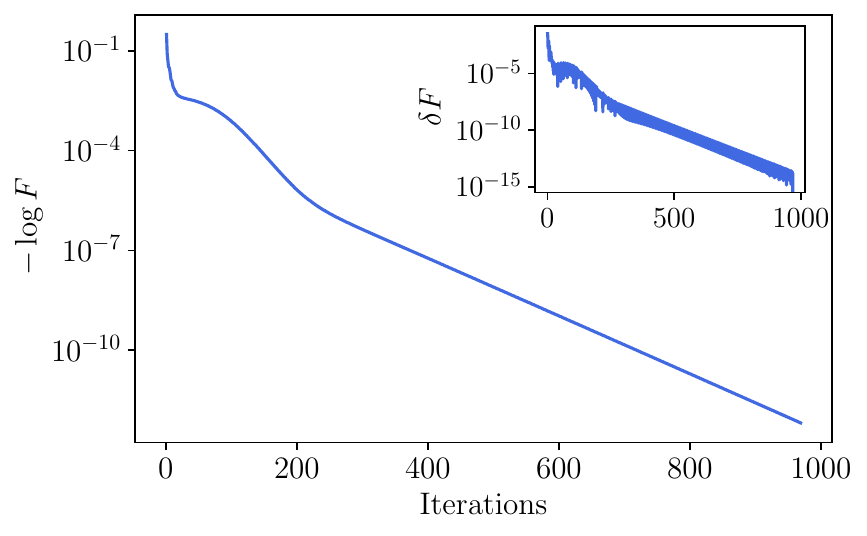}
\caption{Performance of the GD algorithm using the tangent space of iMPS. The algorithm terminates when $\delta F$ reaches $10^{-15}$. The data is for the GS orbit from Fig.~\ref{fig:traj1}. }
\label{fig:conv}
\end{figure}

The convergence of the algorithm for a random initialization is illustrated in Fig.~\ref{fig:conv}. As the algorithm converges, the improvement in fidelity at each iteration decreases until it reaches machine precision, at which point the process terminates. For our search, we set the absolute tolerance (\texttt{atol}) and tolerance (\texttt{tol}) to $10^{-15}$, with the maximal learning rate $max_c = 1$, and parameter $\xi = 1.2$ ($\tau = 1.4$) controlling the dynamical updates decreasing (increasing) the learning rate. The finite difference used to calculate gradients is set to $\Delta = 10^{-5}$, and the learning rate at which the procedure halts is chosen as $\epsilon = 10^{-14}$. The initial value of the learning rate $c$ is chosen as $0.1$. These parameters were selected to ensure both accuracy and computational efficiency. 

Similar to the procedure for finding the canonical form (see the comment at the end of Appendix~\ref{App:canonical}), the gradient descent algorithm also faces a problem when converging to the state that corresponds to a lower bond dimension $\chi'<\chi$ compared to the fixed bond dimension $\chi$. For such instances, we observe that the number of non-zero directions in tangent space decreases from $2\chi^2$ to $2\chi'^2$ at a certain point. We observe that this causes the gradient vector to be ill-behaved: the change of non-zero components of the gradient is very rapid (the second derivative is very large), causing the learning rate of the gradient descent to shrink and leading to the eventual halting of the optimization. 

\section{Classical properties of the periodic orbits}
\label{App:2}
In this appendix, we discuss the classical properties of the periodic orbits. First, we analyze the inversion symmetry. Next, we explain the construction of the Jacobian using a tangent space basis that is required for the study of classical stability of periodic orbits. Using the spectrum of the Jacobian, we also provide details for the onset of stability of the periodic orbit studied in Sec.~\ref{sec:J}. After that, we demonstrate the stable periodic orbit for $\chi = 3$, analyze the Fourier spectra of the correlation functions, and reveal a nine-dimensional KAM-torus. Finally, we compare periodic orbits found for different values of the bond dimension.

\subsection{Inversion symmetry of periodic orbits}
\label{App:inv}
The Kicked Ising model has spatial inversion symmetry, often referred to as parity. However, our algorithm for finding periodic orbits does not impose this symmetry, and we do not restrict our initial guess to have inversion symmetry. Hence, we expect that the orbit search may find both inversion-symmetric and inversion-breaking orbits. Indeed, from the twelve orbits that we discuss in Sec.~\ref{sec:res}, three orbits are not inversion symmetric.
Since the Hamiltonian is inversion-symmetric, all inversion-breaking orbits, have a partner periodic orbit that can be obtained via the action of spatial inversion.  

\begin{figure}[t]
    \centering
    \includegraphics[width=0.98\linewidth]{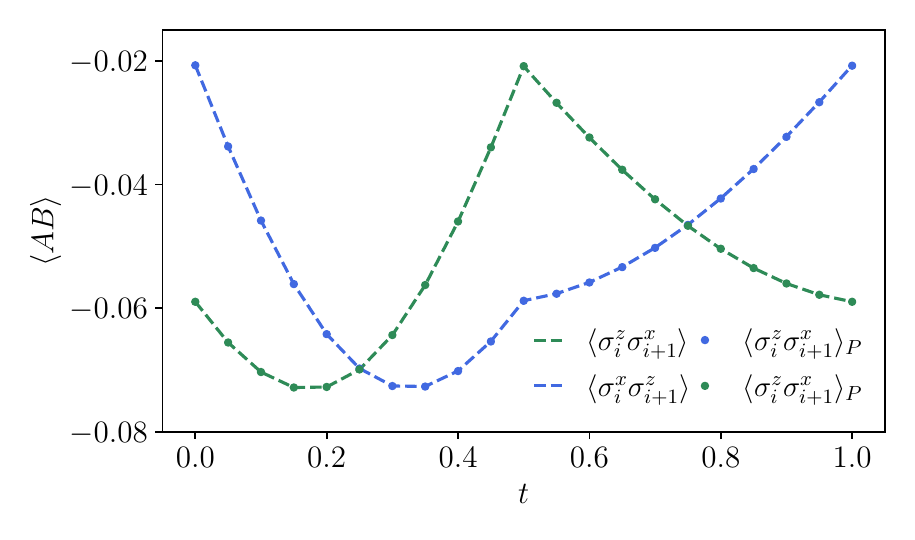}
    \caption{An example of a periodic orbit without inversion symmetry (Fig.~\ref{fig:traj1} $\chi = 3$ green line, 6th from the bottom). The absence of inversion symmetry is manifest in the correlation functions of Pauli operators for adjacent sites, $\langle \sigma^z_i \sigma^x_{i+1} \rangle \neq \langle \sigma^x_{i} \sigma^z_{i+1} \rangle$ shown by green and blue dashed lines. Correlation functions calculated for the spatially inverted periodic orbit (see description in the text) are denoted as $\langle ..\rangle_P$ and are shown by dotted lines.}
    \label{fig:P}
\end{figure}

An example of such a periodic orbit is illustrated in Figure~\ref{fig:P}, where we use the expectation values of Pauli matrices on adjacent sites to demonstrate the absence of inversion symmetry. The same figure also shows expectation values for the spatially inverted orbit. This orbit is obtained by transposing the bond indices of the MPS tensor $A_L(0)$, which correspond to the reflection of spatial coordinates. Such transposed MPS states correspond to different periodic orbits, as shown in Fig.~\ref{fig:P}.

While the occurrence of classical periodic orbits with a lower symmetry compared to the Hamiltonian is natural, it prompts the discussion of the correspondence between low-leakage periodic orbits and eigenstates of the unitary operator generating dynamics over one period, $U_T$. The eigenstates of $U_T$ are expected to obey the same symmetries as the Hamiltonian. Hence, they cannot correspond to inversion-breaking periodic orbits. However, using the fact that each inversion-breaking periodic orbit has a partner with the same period, we can try to form inversion-symmetric and antisymmetric periodic orbits. Notably, these combinations will correspond to a non-injective MPS representation with a larger bond dimension (as a sum or difference of two will have a transfer matrix with two eigenvalues with magnitude one, similar to the GHZ state), that are beyond the applicability of our algorithm. It remains to be seen if the intrinsic nonlinearity introduced by the projection for non-injective MPS may still allow such linear combinations of periodic orbits to remain periodic orbits. In any case, we speculate that the presence of low-leakage inversion-breaking orbits may indicate a tendency to spontaneous inversion symmetry breaking. 

\subsection{Algorithm for the calculation of the Jacobian}
\label{App:Jac}

To construct the Jacobian, we use a tangent space basis, $\{V\}= \{V_\text{re}, V_\text{im}\}$, as it encodes all physical directions corresponding to changes in the quantum state. We first take an initial state corresponding to the periodic orbit at time $t = 0$, $A_L(0)$, and perturb it with strength $\Delta$ along a particular tangent-space direction, $V^{(i)}$. After evolving the perturbed state for one period using TDVP, we expand the resulting state, $A_L^\text{pert}(T)$,  over the tangent basis at the same point $A_L(0)$. Computing the expansion of $A_L^\text{pert}(T)$ over the basis $\{V\}$, we face the problem of the gauge freedom of the MPS. The tangent basis is very sensitive to the gauge, that is why, if during the evolutions state gains some gauge, the difference between $A_L^\text{pert}(T)$ and $A_L(0)$ cannot be expanded over the tangent space basis of $A_L(0)$. 

In order to avoid dealing with gauge fixing, we rely on the same notion of distance between states, Eq.~\eqref{eq:distance}, as used for the search for periodic orbits. Specifically, we choose $\{\beta_i\}$ as a set of parameters for the optimization and construct the tensor $\tilde A_L^\text{pert} = A_L(0)+\beta_i V^{(i)}$, and normalize it. Then, we maximize the fidelity ${\cal D}(A_L^\text{pert}(T), \tilde A_L^\text{pert})$ over the parameters $\{\beta_i\}$ using the \texttt{Optim} package in Julia and taking into account that $\tilde A_L^\text{pert}$ should be always normalized. Effectively this optimization finds the expansion of the perturbation at time $T$ over the tangent space basis using a gauge-invariant cost function. 
The resulting coefficients $\{\beta_i\}$ should then be divided by $\Delta$ to obtain the finite-difference approximation of the given matrix element of the Jacobian, see the expression in Algorithm~\ref{alg:Jac}.

\begin{algorithm}[b]
\caption{Jacobian construction}\label{alg:J}
\label{alg:Jac}
\algorule
\begin{algorithmic}[1]
    \Require $A_L(0)$, $\{V\}$
    \For{$i \in [1, 2\chi^2]$}
    \State $A_L^\text{pert}(0) = A_L(0)+\Delta V^{(i)}$
    \State Normalize $A_L^{pert}(0)$
    \State $A_L^{pert}(T) = U_T(A_L^{pert}(0))$
    \State Introduce parameters $\{\beta_j\}$ and tensor $\tilde A_L^\text{pert} = A_L(0)+\beta_i V^{(i)} $
    \State $\beta_j =  \max_{\{\beta\}} \Big(g(A_L^{pert}(T), \tilde A_L^\text{pert})\Big)$ 
    \State $Jac_{i, j} = \beta_j/\Delta$
    \EndFor
\end{algorithmic}
\algorule
\end{algorithm}

The Jacobian constructed according to Algorithm~\ref{alg:Jac} is influenced by several sources of numerical errors. First, we use a finite difference approximation for the derivatives. For our computations, we choose $\Delta = 10^{-3}$ as we observed that taking smaller values has little effect on the Jacobian matrix. The second source of numerical errors comes from the fact that orbits may not be perfectly periodic due to the imperfect convergence of the gradient descent algorithm. Finally, the third source of errors could potentially come from the numerical maximization of the fidelity ${\cal D}(A_L^\text{pert}(T), \tilde A_L^\text{pert})$. At this point, we encounter the problem when the optimization algorithm selects the correct direction in the space of parameters $\beta$ but results in a slightly larger magnitude of coefficients, as we normalize $\tilde A_L^\text{pert}$ every time we construct it. Normalization is required to compare two MPS states according to Eq.~\eqref{eq:distance}. To avoid this issue, we choose a zero vector as an initial guess for the optimization. In this case, we expect to fall into the local minima near $A_L(0)$. Finally, we expect that the finite-time discretizations of the TDVP equations of motion break the symplectic structure, though it remains to be understood at which order in the time step this occurs. 
As an independent test of the magnitude of numerical errors, we use the product of magnitudes of Jacobian eigenvalues. Due to the preservation of the phase space volume, this is expected to be equal to $1$. For periodic orbits, we typically obtained the product of eigenvalues that differs from one by less than $0.1$ for unstable orbits and less than $0.01$ for stable orbits. 

\subsection{Onset of instability for a $\chi=2$ periodic orbit}
\label{App:stab_J}
\begin{figure}[t]
    \centering
    \includegraphics[width=1\linewidth]{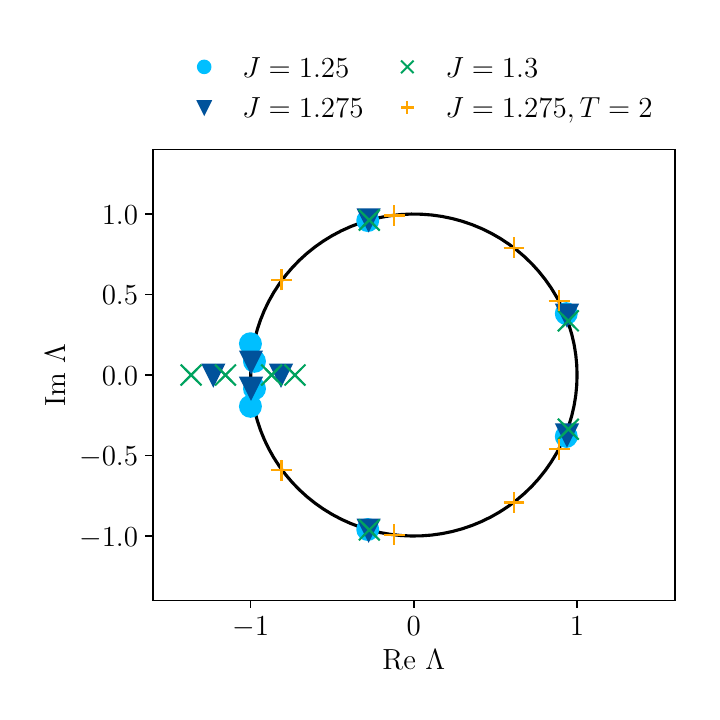}
    \caption{The evolution of Floquet multipliers of the GS orbit in the complex plane upon changing the coupling $J$ in the kicked Ising model. For a stable orbit, all Floquet multipliers lie on the unitary circle. Upon increasing $J$, we observe that two ($J = 1.275$) and then 4 ($J = 1.3$) Floquet multipliers collide and then move away from the unitary circle. This indicates the onset of instability. Orange triangles correspond to the period-doubled stable orbit that we find for the value of coupling past the instability onset.}
    \label{fig:Fm}
\end{figure}
In the main text in Sec.~\ref{sec:J}, we observed that by changing the coupling $J$, the nature of periodic orbit changes from stable to unstable. Here, we study this transition in more detail. In particular, we look at the evolution of all Floquet multipliers (eigenvalues of the Jacobian) in the complex plane (see Fig.~\ref{fig:Fm}). Before instability, for $J = 1.25$, we observe that all $2\chi^2=8$ eigenvalues live on the unit circle in the complex plane, with four of them located near $-1$, and the remaining four distributed over the circle (not visible due to overlap with other dots corresponding to higher $J$). Upon increasing $J$, we notice that the first pair of these 4 nearby eigenvalues is approaching each other. This pair ``collides'' at the point of the instability, and past the onset of instability at $J = 1.275$ one of the eigenvalues becomes larger than $1$ and moves outside of the unit circle, whereas the second eigenvalue moves inside. Even further away, at  $J = 1.3$ we notice that there are already four eigenvalues not on the unit circle.

Observing that a pair of multipliers that were previously on the unit circle are now on the real line and on opposite sides of \(-1\), we assume that \(-1\) was crossed with increasing \(J\), which points at a (supercritical) period-doubling (also known as `flip') bifurcation~\cite{Kuznetsov2023} at some \(J\) between \(J=1.25\) and \(J=1.275\). However, we were unable to find the stable double-period orbit that would correspond to this bifurcation. While for $J = 1.275$ we did find a stable double-period orbit (its stability is recorded in Fig.~\ref{fig:Fm} with orange \(+\) marks), it does not branch from the stable period-one orbit at \(J=1.25\).
\begin{figure}[b]
    \centering
    \includegraphics[width=1\linewidth]{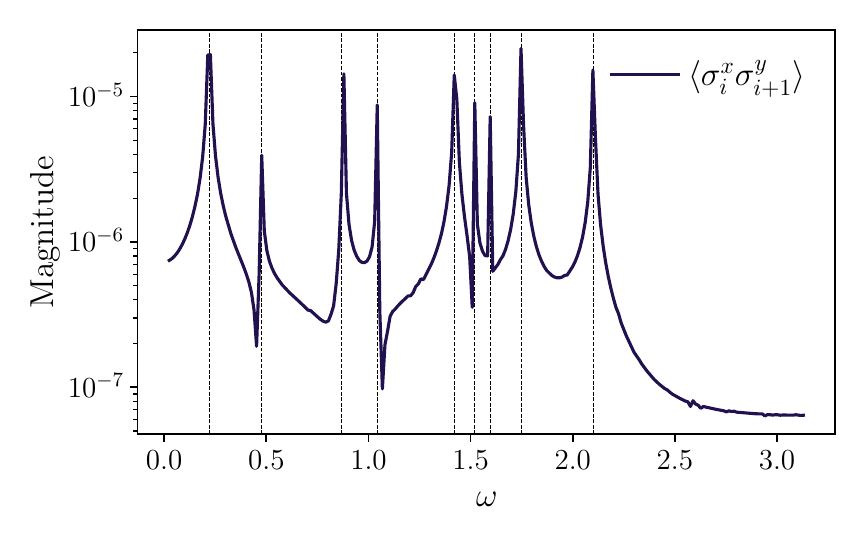}
    \caption{The $\chi = 3$ stable periodic orbit for $J = 0.91$. Frequencies from Jacobian are the same as from Fourier of the signal. The excess entanglement accumulated in the exact dynamics compared to the entanglement on the orbit is $\delta S_{\text{ent}} = 0.11$. Comparing this value with Fig.~\ref{fig:J}, we conclude that this is an intermediate-leakage orbit.}
    \label{fig:FBD3}
\end{figure}

\subsection{Nine-dimensional KAM torus for $\chi=3$}
\label{App:stab_9}
In Sec.~\ref{sec:res}, for the $J = 1.09$, we discuss the properties of the stable orbit with $\chi=2$. In particular, using the dynamics of local observables when the system is initialized slightly away from the periodic orbit, we visualized the four-dimensional KAM torus. The existence of the torus is already apparent from the fact that local observables remain very close to their value at the periodic orbit for long times. Understanding the dimensionality of the torus is more subtle, and we confirmed it by matching the peaks in the Fourier spectrum to the phases of Jacobian eigenvalues. The four-dimensional nature of the torus, identified by the four different frequencies, is consistent with the expectations of having a $\chi^2$ dimensional torus for a generic stable periodic orbit in an MPS manifold with bond dimension $\chi$.

In order to further confirm these expectations, we studied the properties of the periodic orbit with $\chi=3$ found at the value of coupling $J = 0.91$. Perturbing the orbit by $\Delta = 10^{-5}$, we compute the stroboscopic expectation values of observables for times $nT$ with $n \in [1,500]$. The corresponding Fourier spectra are also compared with the phases obtained from the Floquet multipliers. We note that the analysis of single-spin expectation values $\langle \sigma^\alpha \rangle$ with $\alpha=x,y,z$ revealed only eight, instead of the expected nine peaks. However, analyzing more complicated two-site observables such as $\langle \sigma^x_i\sigma_{i+1}^y\rangle$ shown in Fig.~\ref{fig:FBD3} yields the expected $\chi^2=9$ peaks that also match the phases of the Jacobian's eigenvalues. Therefore, we further support our conclusion about the generic occurrence of $\chi^2$-dimensional KAM tori around stable periodic orbits. In addition, we note that with increasing bond dimension, progressively more dynamical variables are involved in describing the correlations of the system, hence suggesting that stability analysis requires input beyond single-spin expectation values. 

\subsection{Comparison of orbits with different bond dimensions}
\label{App:difBD}
In the main text, we illustrated the existence of multiple periodic orbits for different values of the bond dimension $\chi=1,\ldots,4$. This leads to the natural question of whether the orbits found at different values of $\chi$ are related among themselves and potentially also to the eigenstates of the unitary~\eqref{eq:Kick}. To investigate this potential relation, we study the pairwise distances between orbits at $t=0$ using fidelity in Eq.~\eqref{eq:distance}. 

\begin{figure}[t]
    \centering
    \includegraphics[width=0.98\linewidth]{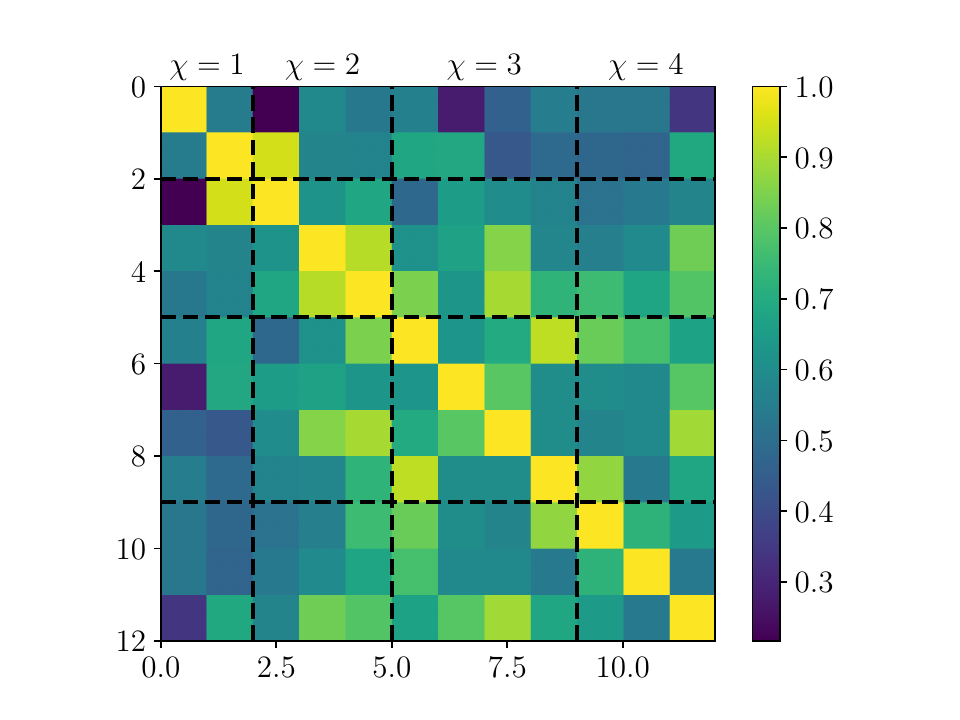}
    \caption{Fidelity between initial points of orbits for different $\chi$. }
    \label{fig:Dist}
\end{figure}

Figure~\ref{fig:Dist} shows the fidelity between the initial points of different orbits. The first periodic orbit $\chi = 1$ is a ceiling state from the main text, and according to the figure, it has relatively small overlaps (at most 0.7) with all remaining orbits. On the other hand, the second orbit found for $\chi = 1$ has a partner $\chi = 2$ orbit, as is witnessed by the considerably larger overlap of $0.95$. This partner orbit was analyzed in Sec.~\ref{sec:J}, where it was shown to be a ground state of the effective Hamiltonian. Note that we did not find further relatives of this orbit at higher bond dimensions (notice the absence of bright colors in the $\chi>2$ parts of the second row or column of the matrix in Fig.~\ref{fig:Dist}). The relation between orbits can be also seen in Fig.~\ref{fig:H_eff} in Appendix~\ref{app:Eff} below, where we illustrate the overlap of the initial points of the orbit with eigenstates of the effective Hamiltonian. 

In contrast to the first $\chi=2$ orbit, the second and third $\chi = 2$ orbits have partners in $\chi = 3$ and $\chi = 4$ as is evidenced by the overlap of $0.86$ and $0.83$ for the second orbit and $0.9$, $0.8$ for the third one. 
Despite the intuition that orbits with higher bond dimensions should better represent the eigenstates of the effective Hamiltonian, in Fig.~\ref{fig:IPR2}, we do not see a qualitative difference in the IPR scaling between orbits with different $\chi$. One reason for this may be the higher leakage of the underlying orbits, as the correspondence between eigenstates and orbits is expected to hold only in the low-leakage regime. Another reason may be related to the algorithm for finding the periodic orbits. As we discussed above in Appendices~\ref{App:canonical} and~\ref{App:alg}, the canonical form and gradient descent algorithms may experience numerical stability issues when converging to the state with an effective lower bond dimension.  

\section{Quantum properties of the periodic orbits} \label{App:quant}
In this Appendix, we provide additional details on the prethermal regime of the kicked Ising model and calculate the overlaps of states corresponding to periodic orbits with eigenstates of the effective Hamiltonian.
\label{App:3}

\subsection{Prethermal regime}
\label{App:Preth}
In this section, we discuss the prethermal regime of the kicked Ising model~\eqref{eq:Kick}. To test at which values of coupling $J$ the system thermalizes, we check if its eigenstates obey Floquet ETH \cite{d2013many, d2016quantum, lazarides2014equilibrium, mori2016rigorous, mori2018thermalization}. For this, we compute the propagator~\eqref{eq:Kick} for $L = 14$ with momentum block $k=0$ and parity block $+1$ for periodic boundary conditions using the $\texttt{QuSpin}$ package \cite{weinberg2017quspin, weinberg2019quspin} and perform exact diagonalization to find eigenstates $|\varphi_n\rangle$,
\begin{equation}
    U_T|\varphi_n\rangle=E_n|\varphi_n\rangle,
    \label{eq:app_fham}
\end{equation}
and corresponding eigenvalues $E_n$.  
The Floquet Hamiltonian relates to the propagator as $H_F=i\ln U_T$, and so the spectrum of the Floquet Hamiltonian is related to that of the propagator
    \begin{equation}
        e_n=i\ln E_n.
        \label{eq:app_eigvals}
    \end{equation}
In the thermal regime, we expect that expectation values of local observables vary only slightly between eigenstates with nearby energies and fluctuate around zero -- their infinite temperature expectation value. In contrast, in the prethermal phase, the eigenstates can be obtained from folding the spectrum of approximate local effective Hamiltonian. Hence, we expect to see the deviations from the typical thermal values.

\begin{figure}[t]
    \centering
    \includegraphics[width=1.\linewidth]{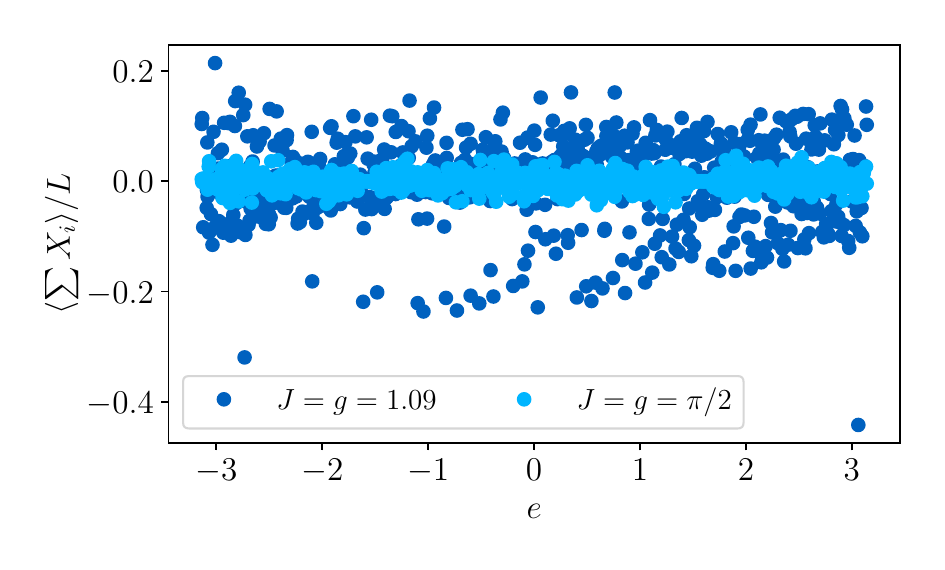}
    \caption{Magnetization along the $x$-direction for different $J$ in a spin chain of size $L = 14$, momentum block 0 and parity block $+1$. For the smaller value of $J$, the systematic dependence on $e$ highlights a prethermal regime, whereas at the dual unitary point, the magnetization fluctuates around zero.}
    \label{fig:Prethermal}
\end{figure}

Figure~\ref{fig:Prethermal} shows a local observable expectation value versus the quasienergy values for two different values of $J$. The behavior at the dual unitary point, $J = \pi/2$, where magnetization fluctuates around zero, should be contrasted to the behavior at the smaller value of $J = 1.09$, where traces of smooth, systematic dependence on the quasienergy are visible, suggesting that system is still in the prethermal regime. The transition from the prethermal regime to the thermal regime leads to the eigenstates of the effective Hamiltonian becoming more and more thermal. This leads to the higher leakage of periodic orbits which we observe in Fig.~\ref{fig:J}(c) top when increasing coupling $J$.

\begin{figure}[b]
    \centering
    \includegraphics[width=1.\linewidth]{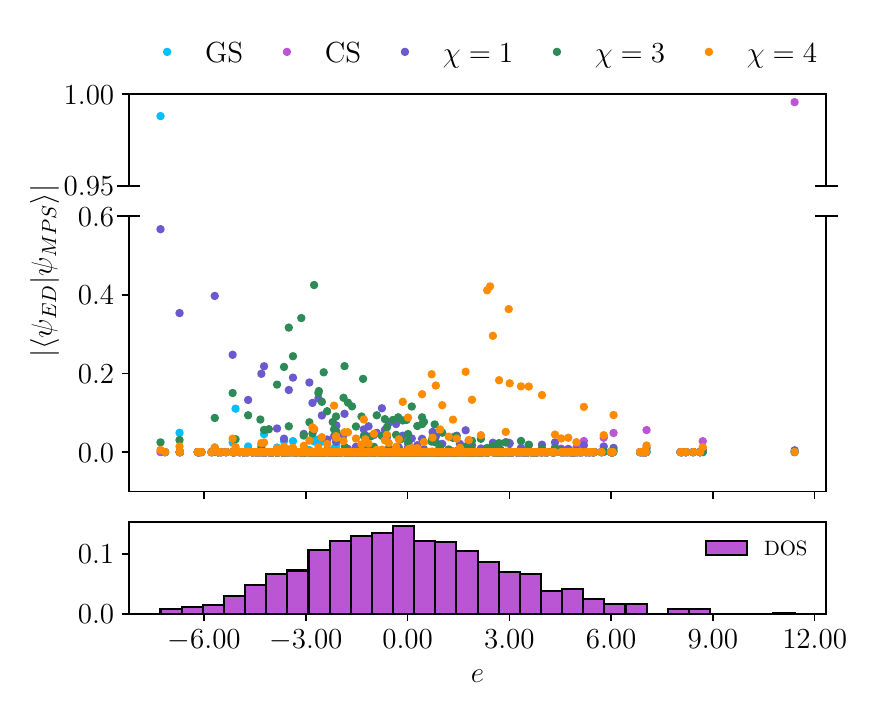}
    \caption{The top panel shows the overlap of some MPS states corresponding to the periodic orbits and eigenstates of 4th order effective Hamiltonian for $J=1.09$ and $L = 10$. The bottom panel shows the normalized density of states (DOS) of the $H_\text{eff}^4$. The different colors of dots correspond to the periodic orbits from Fig.~\ref{fig:traj1} in the main text. The states corresponding to the least leaky periodic orbits have an overlap of order one with the ground and ceiling states of the effective Hamiltonian. One $\chi = 1$ orbit also has a high overlap with the GS, while the $\chi = 3,4$ orbits have the highest overlap with the states from the middle of the spectra.}
    \label{fig:H_eff}
\end{figure}

\subsection{Effective Hamiltonian}
\label{app:Eff}

To understand the role of periodic orbits in the prethermal regime, we study the overlap of the MPS corresponding to the periodic orbit with the eigenstates of the effective Hamiltonian $H_{\text{eff}}^{(4)}$, which is obtained from the fourth-order Magnus expansion of Eq.~\eqref{eq:Kick}. In the kicked model, the Magnus expansion simplifies to the Baker–Campbell–Hausdorff formula: 
\begin{equation}
    \begin{gathered}
        U_T = \text{exp}\Big(-i\frac{1}{2}H_2\Big)\text{exp}\Big(-i \frac{1}{2}H_1\Big) \approx \text{exp}(-iTH^{(4)}_{\text{eff}}), \\
        H_{\text{eff}}^{(4)} = \frac{1}{2}(H_1 + H_2) - \frac{i}{8}[H_2, H_1] + \frac{1}{96}([H_1,[H_1, H_2]] +\\ [H_2,[H_2,H_1]]) - \frac{i}{384}([H_2,[H_1,[H_1, H_2]]]]),
    \end{gathered}
    \label{eq:eff}
\end{equation}  
leading to a translationally invariant Hamiltonian with terms including products of up to five Pauli matrices.

We construct the effective Hamiltonian $H_{\text{eff}}^{(4)}$ using the \texttt{QuSpin} \cite{weinberg2017quspin, weinberg2019quspin} package for $L = 10$ in the full Hilbert space and diagonalize it to obtain all energies ($e_n$) and eigenvectors. 
Figure~\ref{fig:H_eff} shows the overlap of the eigenvectors of the effective Hamiltonian \eqref{eq:eff} with the MPS states corresponding to some of the periodic orbits from Fig.~\ref{fig:traj1}. The two low leakage periodic orbits (GS and CS in Fig.~\ref{fig:traj1}) correspond to the ground and ceiling states of the effective Hamiltonian, as is witnessed by their overlaps being close to one in Fig.~\ref{fig:H_eff}. We also illustrate another $\chi = 1$ orbit, as it has a high overlap with the ground state. 

In contrast, the other MPS states corresponding to more leaky periodic orbits have largest overlaps with eigenstates of the effective Hamiltonian from the middle of the spectrum (For clarity, only two such orbits labeled as $\chi=3,4$ are shown in Fig.~\ref{fig:H_eff}). These overlaps have a maximum at a certain energy and decay exponentially with energy difference (not shown). This is in contrast to the DMRG approximation of ground states of local Hamiltonians, where overlap was found to have flat tails \cite{silvester2025unusual}, and resembles the energy distribution of states obtained from imaginary time evolution. Since these MPS states have a low bond dimension, $\chi\leq 4$, the fact that they have overlaps on the order of $0.5$ with individual eigenstates is still highly unusual. This suggests that even the effective Hamiltonian~\eqref{eq:eff} may be not fully thermal, at least for the considered system size. We defer the systematic investigation of the ETH for the effective Hamiltonian and potential relation of periodic orbits to quantum many-body scars \cite{Turner2017,Serbyn:2021vc,Moudgalya_2022,chandran23,HummelPRL,Dag24PRL,ermakov2024,pizzi2024,Dag24,LjubotinaPRXQ} to future work. 
        
\bibliography{bib}
\end{document}